\documentclass[10pt,journal]{IEEEtran}
\usepackage{amsmath,amsfonts}
\usepackage{algorithmic}
\usepackage{algorithm}
\usepackage{array}
\usepackage[caption=false,font=normalsize,labelfont=sf,textfont=sf]{subfig}
\usepackage{textcomp}
\usepackage{stfloats}
\usepackage{url}
\usepackage{verbatim}
\usepackage{graphicx}
\usepackage{cite}
\usepackage{siunitx}
\usepackage{cleveref}
\usepackage{xcolor}
\usepackage[acronym]{glossaries}
\usepackage{graphicx}
\usepackage{amssymb}
\usepackage{array}
\usepackage{tikz}
\usepackage{pgfplots}
\pgfplotsset{compat=1.18}
\usepgfplotslibrary{units}
\usepackage{booktabs}

\usetikzlibrary{fit}

\usepgfplotslibrary{external}
\tikzset{
  external/system call={pdflatex \tikzexternalcheckshellescape --halt-on-error --interaction=batchmode --jobname "\image" "\texsource"}
}
\tikzexternaldisable %

\newenvironment{externalize}[1]{
\tikzexternalenable %
\tikzsetnextfilename{img-#1} %
}{}

\pgfplotsset{
            /pgfplots/layers/niceLayers/.define layer set={
                        axis background,
                        axis grid,
                        main,
                        axis ticks,
                        axis lines,
                        axis tick labels,
                        axis descriptions,
                        axis foreground
            }{/pgfplots/layers/standard}
}

\pgfplotsset{
            every axis/.append style={
                        set layers=niceLayers,
                        tick label style={font=\scriptsize},
                        clip marker paths=true,
                        line width=1pt,
                        line cap=round,
                        line join=round,
                        tick style={semithick, color=black},
                        legend style={
                        		   font=\scriptsize,
                                    /tikz/every even column/.append style={column sep=2mm},
                                    cells={anchor=west}, %
                        },
                        xmajorgrids,
                        ymajorgrids,
            }
}

\pgfkeys{/pgf/number format/.cd,1000 sep={}}

\newacronym{pu}{PU}{power unit}
\newacronym{f1}{F1}{Formula 1}
\newacronym{cfd}{CFD}{computational fluid dynamics}
\newacronym{mpc}{MPC}{model predictive control}
\newacronym{drs}{DRS}{Drag Reduction System}
\newacronym{mguk}{MGU-K}{motor-generator unit -- kinetic}
\newacronym{nn}{NN}{neural network}
\newacronym[plural=OCPs, firstplural=optimal control problems]{ocp}{OCP}{optimal control problem}
\newacronym{nlp}{NLP}{nonlinear program}
\newacronym{kkt}{KKT}{Karush-Kuhn-Tucker}
\newacronym{mpcc}{MPCC}{mathematical program with complementarity constraints}
\newacronym{licq}{LICQ}{linear independence constraint qualification}
\newacronym{mfcq}{MFCQ}{Mangasarian-Fromovitz constraints qualification}
\newacronym{svo}{SVO}{Social Value Orientation}

\newtheorem{Ass}{Assumption}

\newtheorem{problem}{Problem}

\newcommand{\BCvector}{\mathbf{p}_\mathrm{bc}}

\newlength{\myfigskip}
\begin{document}

\title{Game-theoretic Energy Management Strategies \\ With Interacting Agents in Formula 1}

\author{Giona Fieni, Marc-Philippe Neumann, Alessandro Zanardi, Alberto Cerofolini, Christopher H. Onder~\IEEEmembership{}
\thanks{G. Fieni, M.-P. Neumann, A. Zanardi, C. H. Onder are with the Institute of Dynamic Systems and Control, ETH Zürich, Sonneggstrasse 3, 8092 Zürich, Switzerland (e-mail: \{gfieni, mneumann, azanardi, onder\}@idsc.mavt.ethz.ch). A. Cerofolini is with the Power Unit Performance Group, Ferrari S.p.A., 41053 Maranello, Italy (e-mail: Alberto.Cerofolini@ferrari.com).}}%

\IEEEpubid{\begin{minipage}{\textwidth}\ \centering \\[10pt]
		\copyright~2025 IEEE.  Personal use of this material is permitted.  Permission from IEEE must be obtained for all other uses, in any current or future media, including reprinting/republishing this material for advertising or promotional purposes, creating new collective works, for resale or redistribution to servers or lists, or reuse of any copyrighted component of this work in other works.
\end{minipage}}

\maketitle

\begin{abstract}
This paper presents an interaction-aware energy management optimization framework for Formula 1 racing. The scenario considered involves two agents and a drag reduction model. Strategic interactions between the agents are captured by a Stackelberg game in the form of a bilevel program. To address the computational challenges associated with bilevel optimization, the problem is reformulated as a single-level nonlinear program employing the Karush-Kuhn-Tucker conditions. The proposed framework contributes towards the development of new energy management and allocation strategies, caused by the presence of another agent. For instance, it provides valuable insights on how to redistribute the energy in order to optimally exploit the wake effect, showcasing a notable difference with the behavior studied in previous works. Robust energy allocations can be identified to reduce the lap time loss associated with unexpected choices of the other agent. It allows to recognize the boundary conditions for the interaction to become relevant, impacting the system's behavior, and to assess if overtaking is possible and beneficial. Overall, the framework provides a comprehensive approach for a two-agent Formula 1 racing problem with strategic interactions, offering physically intuitive and practical results. 
\end{abstract}


\section{Introduction}
\IEEEPARstart{F}{ormula 1} consists of 20 pilots racing on a circuit for a predefined number of laps, with the goal to cross the finish line first. Only one driver can claim the win, but the others aim anyway for the best possible placement, since points are distributed for the driver's and constructor's championship. This competitive spirit drives the intensity of \gls{f1} races, with every participant pushing themselves and their vehicle to the limit. Not only the pilot is responsible for the performance: The car developed by the team must be reliable and fast. To achieve the success aimed for, \gls{f1} manufacturers must rely on the forefront of innovation, pushing the boundaries. Similar to other sports, \gls{f1} is also subject to technical and sporting regulations \cite{2024F1, 2024F1_sport}. For this reason, the teams have to exploit each possible opportunity to enhance the performance of the car and gain advantage over others, from the aerodynamics to the control algorithms.

\begin{figure}
\centering
\includegraphics[width=0.85\columnwidth]{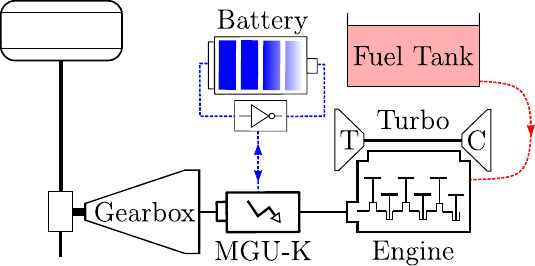}
\caption{Schematic of the \gls{f1} \gls{pu}. The prime movers of the hybrid-electric configuration are the turbocharged internal combustion engine and the \gls{mguk}. The battery and the fuel tank are the onboard energy storages. Through the gearbox, the power coming from the \gls{pu} is transferred to the wheels.}
\label{fig:F1PU}
\end{figure}

Since 2014, \gls{f1} has moved to a hybrid-electric configuration. A sketch of the \gls{pu} is depicted in \Cref{fig:F1PU}. The onboard energy storages are the fuel tank and the battery. The former feeds a V6 $\qty{1.6}{\liter}$ turbocharged engine, while the latter is coupled to an electric motor, the so-called \gls{mguk}. This electrical machine can drain energy from the battery or recharge it during the braking phases. Since refueling is not allowed, the energy for the entire race is limited, raising the need for an energy management. The complex topology renders the research towards an optimal operation challenging. The energy management is influenced by many factors, and in this paper we aim to study the impact of the interactions between racing cars.  

\IEEEpubidadjcol

\subsection{Relevant Interactions}
\IEEEpubidadjcol
During racing, an \gls{f1} car mainly experiences two forces arising from the interaction between its body and the aerodynamic flow: the drag resistance force and the downforce. The former acts against the forward movement, whilst the other literally pushes the car into the ground, affecting the maximal achievable longitudinal and lateral accelerations. Both forces are strongly dependent on the aerodynamic configuration and properties of each car, which may favor one or the other. To maximize the performance, it would be ideal to achieve the lowest drag and the highest downforce. However, the two goals are conflicting: The downforce is generated through aerodynamic devices such as front and rear wing, but each of them increases the drag resistance \cite{milliken1995race}. 

The effects of drag and downforce are relevant in different sections of the track. The drag force scales with the square of the velocity, thus its effect is more prominent on the straights, where the velocity is high. The downforce significantly enhances the grip of the car, but this can be exploited only in the corners where, in comparison to the straights, the peak velocities are considerably lower.

The aerodynamic flow pattern strongly affects the magnitude of these forces. A car alone on the track is said to be in free stream, since it goes through an unperturbed flow. Under these circumstances, drag and downforce are purely generated by the car's velocity. The situation is different behind the car, because the flow remains perturbed and does not return to the initial conditions instantaneously. So, if another car is following closely, the experienced flow conditions are not the same as they would be in free stream. As a matter of fact, drag- and downforce parameters for an \gls{f1} car could be reduced by $\qty{30}{\percent}$, respectively, $\qty{50}{\percent}$ \cite{ravelli2021aerodynamics}, although they strongly vary, based on car and year (more on this in \Cref{chp:litRev}). 

The impact of these interaction effects is highly relevant for the overall performance of the racing car. Whilst the associated drag reduction is an advantage, since it dissipates less energy, the downforce reduction has the downside of decreasing the maximal cornering velocity. The magnitude of the loss is mainly determined by the distance between the cars and by their velocity.

The \gls{drs} is another relevant topic when considering interactions between two Formula 1 cars. The aerodynamics of the car can be actively changed to a lower drag configuration by flattening the rear wing. This system is meant to help overtaking by quickly reducing the gap time. However, it is only available if the measured gap time at a specific location is below $\qty{1}{\second}$. Since the decrease in drag comes along with reduced downforce, it is allowed only on predefined zones on the track for safety considerations.

The impact on the energy management of these three factors is non-trivial, being linked also to strategic choices such as overtakes. Provided that another car is at a tangible distance such that the drag reduction becomes prominent, the own drag can be reduced. This allows the car behind to be faster or to save energy with a view to a future overtake. On the other hand, the downforce loss decreases the grip of the car, affecting the distance to hold during cornering. The \gls{drs} adds a level of complexity, since one could aim to reduce the gap time before the \gls{drs} detection zones to gain the advantage of its usage. The energy management strategy can therefore be adapted to meet these goals, and quantifying the impacts is crucial.

The focus of this paper is to take a first step in the direction of an optimal energy management, by considering the active response from other agents. To simplify the task, we decided not to consider the \gls{drs}, since given its nature it is likely to introduce integer decision variables and non-smooth dynamics in the optimization. Furthermore, while it would be possible to extend the presented approach to include both drag and downforce reduction, we limit the analysis to the drag reduction. Having only one influence factor simplifies the interpretation of the results for a dynamic environment with two agents. Furthermore, the drag reduction has a direct impact on the energy management, the peak velocities and the overtakes, covering a considerable amount of typical situations arising during an \gls{f1} race. Although there exists some overlapping with the downforce reduction, their relevance is found in different sections of the track (straights or corners), allowing for a separation of the analysis.

The developed optimization framework serves as a robust basis for future analyses and more complex modeling.

\subsection{Literature Review}\label{chp:litRev}
The respective research areas relevant for this paper can be divided into three categories.\\
The first one regards the aerodynamic interaction of two vehicles following each other closely, with a focus on the wake effect. Although the main application field is on racing cars, where the effect is more prominent, there are studies such as \cite{hong1998drag}, which experimentally tests platooning in road vehicles to enhance energy efficiency. Regarding \gls{f1} or general open-wheeled racing cars, we find extensive \gls{cfd} studies in \cite{ravelli2021aerodynamics,mafi2007investigation,guerrero2020aerodynamic,newbon2015cfd}, experiments in \cite{dominy1990influence,soso2006aerodynamics} and a combination of the two in \cite{newbon2017aerodynamic, newbon2017aerodynamic2}. Their common points are the following: They consider cars with the same velocity, and the impact that the wake has on the drag and downforce is expressed as a function of the longitudinal spacing between them. Although the results are not directly comparable due to structural differences of the models, they all show a similar trend in the magnitude of the drag reduction of the subsequent vehicle. For \gls{f1} cars, the maximum reduction ranges from around $\qty{40}{\percent}$ \cite{guerrero2020aerodynamic} to $\qty{22}{\percent}$ \cite{dominy1990influence}. Similar results were also found in \cite{dvzijan2021aerodynamic} and \cite{gan2020cfd} for closed-wheel racing cars, such as NASCAR. In particular, in \cite{gan2020cfd} the author considers a leading car and up to three trailing cars in close proximity. Additionally, \cite{newbon2017aerodynamic} and \cite{gan2020cfd} study the influence on the drag and downforce given from the lateral shift. Despite extensive investigations, integrating these simulations in a dynamic optimization framework remains unresolved, with a notable absence of literature in this direction. Furthermore, the influence of the wake effect on the energy management represents another unexplored topic in the field. 

The second area covers the topics of modern hybrid-electric racing cars, with the focus on energy management or the presence of competitors. In the field of \gls{f1}, considerable research effort is invested in convex \cite{ebbesen2017time,duhr2022convex,balerna2020optimal} and nonlinear \cite{balerna2020time} lap time optimizations. Results show the strong coupling between energy management and \gls{pu} operation, for instance the influence on the power split or on the gearshift strategy. Additionally, they serve as a starting point to develop control strategies. The authors in \cite{salazar2017time} derive an analytical optimal control policy, whereas in \cite{salazar2017real} they employ a convex two-level \gls{mpc} scheme. In \cite{neumann2023low}, a nonlinear \gls{mpc}, responsible for the low-level operation of the \gls{pu}, is coupled to the analytical optimal control policy, which takes care of the energy management. Energy allocation strategies for the entire race are discussed in \cite{duhr2023minimum}. The comparison between an optimal allocation of fuel and battery energy with respect to a heuristic one showed an improvement of \qty{2}{\second} of the \textit{race} time. Concerning the inclusion of competitors, \cite{paparusso2022competitors} proposes an endurance racing model featuring an ego-vehicle aware of the presence of competitors. Using a statistical approach, the author characterizes sectors with probabilities of overtaking different categories of vehicles. An optimization of the ego-vehicle allows to identify the best strategy at the beginning of each lap to improve the lap time in the presence of traffic. Energy efficient overtaking is investigated in \cite{liu2023energy} for Formula E cars. The model includes longitudinal and lateral dynamics together with avoidance constraints and the ego-vehicle optimizes its trajectory given the pre-computed reference trajectory of the target car. The literature gap in this direction is given by the following facts: The impact of competitors on the energy management is either not included or not investigated through aerodynamic interaction, although this type of physical interaction is often exploited. In addition, the active response of other agents is neither considered nor modeled. 

The third relevant field is the use of game theory for racing applications and vehicles control. Autonomous racing is a widely studied topic in combination with game-theoretic approaches, in drone racing \cite{spica2020real,wang2019game2,wang2020multi}, car racing \cite{liniger2019noncooperative,notomista2020enhancing,wang2019game,wang2021game,williams2018best} or even sailboats competitions \cite{cacace2020stochastic}. The focus of these works lies on trajectory planning, collision avoidance and estimation of opponents' behavior. A popular approach is to combine a best-response algorithm with a receding horizon control for all the involved agents. We can find different versions, ranging from Nash seeking best-response algorithms \cite{wang2019game,wang2019game2,wang2020multi,wang2021game}, to algorithms with an augmented sensitivity term to account for collisions \cite{notomista2020enhancing,spica2020real}, to stochastic \gls{mpc} combined with a best-response algorithm \cite{williams2018best}. In the presented applications, the game-theoretic planners performed better than classical \gls{mpc}. In the sailboats competition of \cite{cacace2020stochastic}, a different type of approach is employed. Value iteration is combined with a numerical scheme to approximate the solution of the Hamilton-Jacobi-Bellmann-Isaac equation of the game. Besides the strategic route planning, the author mentions that the problem could have potentially included the wind shadow effect, which was neglected. This further justifies the novelty of a drag interaction model in an optimization problem with multiple agents. For road vehicles, \cite{burger2022interaction} proposes a lane changing motion planning algorithm for an autonomous vehicle interacting with a human driver. Cooperative behavior is introduced using courtesy constraints or through additional terms in the cost function. The problem is solved in receding horizon using a Stackelberg game formulation. In contrast to \cite{paparusso2022competitors,liu2023energy}, this research area focuses on the active response of agents, rather than just accounting for their presence. The most popular approach is to solve these problems in a receding horizon fashion, solving relatively small problems. In conclusion, in the field of \gls{f1}, neither game-theoretic approaches for energy management strategies are employed, nor are other physical interactions investigated such as drag or downforce reduction.

\subsection{Research Statement}
The aim of the presented work is to bridge the literature gap between optimal energy management and interaction between agents. In contrast to works which consider interactions such as collision avoidance, we focus on the drag reduction arising from the presence of another agent. 

The main body of literature on multi-agent interactions in robotics focuses on the motion planning aspects, by predicting others' behaviors, reactions, and computing a collision-free trajectory. Game-theoretic approaches in racing scenarios mainly consider only trajectory planning for overtaking and blocking manoeuvers in a receding horizon fashion.
Conversely, in \gls{f1} racing, aerodynamic effects play a crucial role in the energy management. We research the question of how to strategically optimize the energy management system of a racing car in the presence of another agent.

Furthermore, we aim to investigate the possible differences in energy management strategies compared to prior studies \cite{balerna2020time,duhr2022convex,duhr2023minimum,ebbesen2017time}. To the best of our knowledge, a drag reduction model has not yet been included in previous optimization frameworks, nor its impact on the energy management of both agents investigated. 

\subsection{Contributions}
The contribution of our work is threefold: First, we propose a computationally efficient optimization framework which includes a drag reduction model to consider its entanglement with the energy management.

Second, we introduce strategic responses between the agents using a Stackelberg game formulation for the \gls{f1} lap problem. The game-theoretic approach captures the interaction-aware decision-making process. 
To the best of our knowledge, the interaction between \gls{f1} cars and their energy management has never been included in a dynamic game to date.

Third, we showcase the impact of the physical interaction on the energy management and on the lap time. We analyze the optimal redistribution of the saved energy and we investigate robust energy allocation strategies based on the other agent's choice. The presented optimization framework can handle a large variety of initial conditions, and it lays the basis to develop new strategies to improve the lap time.

Although the case study considered for this work is a \gls{f1} race, this approach is also applicable to other highly dynamic fields, such as endurance competitions or Formula E races. Given their duration, the energy savings potential due to aerodynamic interactions are considerable.

\subsection{Outline}
This paper is structured as follows: In \Cref{sec:modelinter}, we present the models of the agents, together with the drag reduction model which describes their interaction. Next, we describe in detail the game-theoretic approach for the problem formulation in \Cref{sec:gametheory}, and the results are discussed in \Cref{sec:results}. Finally, we conclude the paper in \Cref{sec:conclusion} with an outlook on potential extensions of the presented framework.

\section{Agents: Modeling and Interaction}\label{sec:modelinter}
To study the interaction mentioned above, we consider two agents $A$ and $B$. To distinguish between them, we use the subscript $i$, where $i\in\{A,B\}$. After describing their dynamic model, we introduce the drag reduction model which captures the interaction. The characteristics of each car are different, but they are governed by the same equations. For the sake of simplicity, we consider two identical cars. The physical parameters can be easily changed within the framework, enabling the possibility to analyze cars of different teams, or to adapt to the updates that take place during the season. Finally, their \glspl{ocp} are summarized. 

\subsection{Model of a Single Agent}
Here, we present the dynamic model of the single agent, and to improve the readability, the subscript $i$ is dropped. Similarly to \cite{ebbesen2017time}, we formulate the problem in the space domain, because the track properties are space-based. We make use of the distance variable $s\in [0,S]$, where $S$ corresponds to the track length. To switch from time to space domain, analogous to \cite{balerna2020time} we exploit the definition of velocity to employ the following change of variables:
\begin{equation}\label{dsdt}
v(t) = \frac{\mathrm{d}s(t)}{\mathrm{d} t} \quad \leftrightarrow \quad \mathrm{d}t(s) = \frac{\mathrm{d}s}{v(s)}.
\end{equation}

We start now by describing the powertrain model, whose topology is depicted in \Cref{fig:F1PU}. The inputs of each agent are its \gls{mguk} power $P_\mathrm{k}$, its fuel power $P_\mathrm{f}$, and the friction brake power $P_\mathrm{brk}$. The physical limits on the \gls{mguk} power are defined by the sporting regulations \cite{2024F1}, as well as the maximal fuel mass flow $\dot{m}_\mathrm{f,max}$. These physical limits result in the input space
\begin{align}
P_\mathrm{k,min} \le & P_\mathrm{k}(s) \le P_\mathrm{k,max} \label{eq:pkbound}\\
0 \le & P_\mathrm{f}(s) \le \dot{m}_\mathrm{f,max} \cdot H_\mathrm{l} \label{eq:pfbound}\\
0 \le & P_\mathrm{brk}(s),  \label{eq:pbrkbound}
\end{align}
where $H_\mathrm{l}$ is the lower heating value of the fuel. Since the \gls{mguk} is used in both motor and generator mode, it holds that $P_\mathrm{k,min}\le 0$ and $P_\mathrm{k,max}\ge 0$. To model the engine power, we make use of the Willans model \cite{guzzella2009introduction} 
\begin{equation} 
P_\mathrm{e}(s) = \eta_\mathrm{e}\cdot P_\mathrm{f}(s) - P_\mathrm{e,0},  \label{eq:pengine}
\end{equation}
where $\eta_\mathrm{e}$ represents the Willans efficiency and $P_\mathrm{e,0}$ the engine drag power. 
This simple model is precise enough to link the engine power with real fuel consumption values. Widely used in previous works \cite{ebbesen2017time}, it has proven effective for the considered purpose and timescale. We can define the propulsive power $P_\mathrm{p}$ coming from the \gls{pu} as
\begin{equation}
P_\mathrm{p}(s) = P_\mathrm{e}(s) + P_\mathrm{k}(s) - P_\mathrm{brk}(s),  \label{eq:propulsive}
\end{equation}
where we assume a perfect transmission efficiency.

Next, we model the agent's states, which are relevant to describe the interplay between the performance, the energy management and the drag interaction. The energy storages are the kinetic energy (represented by the velocity $v$), the consumed fuel energy $E_\mathrm{f}$, and the battery energy $E_\mathrm{b}$. The time $t$ serves as information storage, and plays a crucial role in our model for the characterization of the interaction forces. Given the change of variables of \eqref{dsdt}, the differential equation of the time in space domain reads
\begin{equation}
\frac{\mathrm{d}}{\mathrm{d} s} t(s)  = \frac{1}{v(s)}. \label{eq:time}
\end{equation}
Whilst the final time is subject to optimization, the initial time $t_\mathrm{init}$ is given as boundary condition 
\begin{equation}
 t(0) = t_\mathrm{init}. \label{eq:tinit}
\end{equation}
With this definition, we can change the initial distance by providing each agent a different $ t_{\mathrm{init},i}$. The dynamics of the fuel energy are described by
\begin{equation}
\frac{\mathrm{d}}{\mathrm{d} s} E_\mathrm{f}(s)  = \frac{P_\mathrm{f}(s)}{v(s)}. \label{eq:fuel}
\end{equation}
Since the fuel power $P_\mathrm{f}$ cannot take negative values, the consumed fuel energy $E_\mathrm{f}$ can only increase. Its boundary conditions read 
\begin{equation}
 E_\mathrm{f}(0) = 0, \qquad E_\mathrm{f}(S) \le E_\mathrm{f,target}, \label{eq:efbc}
\end{equation}
with $E_\mathrm{f,target}$ the allocated fuel energy for the current lap. The battery dynamics evolve as 
\begin{equation}
\frac{\mathrm{d}}{\mathrm{d} s} E_\mathrm{b}(s)  = -\frac{P_\mathrm{k}(s)}{v(s)}. \label{eq:battery}
\end{equation}
When using the \gls{mguk} in motor mode, $P_\mathrm{k}$ is positive and the battery is discharged. Conversely, when using the \gls{mguk} in generator mode, the battery is charged. To simplify, we neglect any losses taking place during energy conversions (internal battery losses, electrical-to-mechanical and vice versa). For more details on this, the reader is referred to \cite{ebbesen2017time}. The initial battery charge $E_\mathrm{b,init}$ and the allocated difference in battery energy $\Delta E_\mathrm{b}$ defines the boundary conditions for this state:
\begin{equation}
 E_\mathrm{b}(0) = E_\mathrm{b,init}, \qquad E_\mathrm{b}(S) \ge E_\mathrm{b,init} + \Delta E_\mathrm{b}. \label{eq:ebbc}
\end{equation}
The formulation of the boundary conditions for fuel and battery energy is the same as in \cite{duhr2022convex} and they are stated as inequality constraints. Therefore, the agent is not forced to use all the energy and possibly employ suboptimal strategies to get rid of the energy surplus, as shown in \cite{duhr2023minimum}. In addition, it increases the feasible set of the problem. Furthermore, the battery has a finite capacity which results in the constraint
\begin{equation}
 0\le E_\mathrm{b}(s) \le E_\mathrm{b,max}. \label{eq:ebbounds}
\end{equation}
The car's kinetic energy $E_\mathrm{kin}$ and velocity are linked by the relation
\begin{equation}\label{eKinV}
 E_\mathrm{kin} = \frac{1}{2}\cdot m\cdot v^{2},
\end{equation}
where $m$ is the mass of the car, assumed constant over the lap. Using the definition of kinetic energy, we can characterize the longitudinal dynamics in time domain as
\begin{equation}\label{eKinTime}
\frac{\mathrm{d}}{\mathrm{d} t} E_\mathrm{kin}(t) = P_\mathrm{p}(t) - P_\mathrm{ext}(t)
\end{equation}
where $P_\mathrm{ext}$ describes the power associated to the external forces. Since \eqref{eKinV} is valid in both space and time domain, the left hand side of \eqref{eKinTime} can be stated as 
\begin{equation}
\frac{\mathrm{d}}{\mathrm{d} t} E_\mathrm{kin}(t) =\frac{1}{2} m\cdot \frac{\mathrm{d}}{\mathrm{d} t} v^{2}(t),
\end{equation}
and by applying the chain rule we get 
\begin{equation}
\frac{\mathrm{d}}{\mathrm{d} t} E_\mathrm{kin}(t) =m\cdot v(t)\cdot \frac{\mathrm{d}}{\mathrm{d} t} v(t).
\end{equation}
Applying the change of variables \eqref{dsdt}, the spatial derivative becomes
\begin{equation}
v(s)\cdot \frac{\mathrm{d}}{\mathrm{d} s} E_\mathrm{kin}(s) = m\cdot v^{2}(s)\cdot \frac{\mathrm{d}}{\mathrm{d} s} v(s). \label{eKinSpace}
\end{equation}
Combining \eqref{eKinTime} and \eqref{eKinSpace}, the resulting dynamics for the velocity in space domain are  
\begin{equation}
\frac{\mathrm{d}}{\mathrm{d} s} v(s) = \frac{P_\mathrm{p}(s) - P_\mathrm{ext}(s)}{m\cdot v^{2}(s)}.\label{eq:velocity}
\end{equation}
The initial velocity is chosen as a boundary condition, and the final velocity is subject to optimization. Furthermore, we implicitly embed the lateral dynamics of the car by means of a space-dependent maximum velocity profile $v_\mathrm{max}$. The industrial partner derived it from measurements and simulations, and it assumes a single racing line. These limits result in the constraints
\begin{equation}
v(0) = v_\mathrm{init}, \qquad v(s) \le v_\mathrm{max}(s). \label{eq:vbc}
\end{equation}

In the following, we characterize the power associated to the external forces $P_\mathrm{ext}$. Force and power are linked by  
\begin{equation}
P_\mathrm{ext}(s) = F_\mathrm{ext}(s)\cdot v(s),  \label{eq:pext}
\end{equation}
and the contributions of each external force is given by the sum of the total aerodynamic drag $F_\mathrm{aero,tot}$, the rolling resistance $F_\mathrm{roll}$, and the projected weight force $F_\mathrm{slope}$ stemming from the track's slope:
\begin{equation}
F_\mathrm{ext}(s) = F_\mathrm{aero,tot}(s) + F_\mathrm{roll} + F_\mathrm{slope}(s).  \label{eq:fext}
\end{equation}
The rolling resistance $F_\mathrm{roll}$ is assumed to be constant and proportional to the car's weight
 \begin{equation}
F_\mathrm{roll} = c_\mathrm{roll} \cdot m\cdot g,  \label{eq:froll}
\end{equation}
with $c_\mathrm{roll}$ the rolling resistance's coefficient and $g$ the gravitational acceleration. The last term of the external forces acting on the car is the weight force of the car projected on the direction of motion, characterized by the slope of the track $\alpha$:
 \begin{equation}
F_\mathrm{slope}(s) = m\cdot g\cdot \sin\left(\alpha(s)\right).  \label{eq:fslope}
\end{equation}
The total aerodynamic drag $F_\mathrm{aero,tot}$ is composed of 
\begin{equation}
F_\mathrm{aero,tot}(s) = F_\mathrm{aero,fs}(s) - F_\mathrm{aero,int}(s), \label{eq:faerotot}
\end{equation}
where $F_\mathrm{aero,fs}$ is the drag force as if the agent were in free stream, and $F_\mathrm{aero,int}$ is the reduction coming from the interaction with the other agent. The drag force is proportional to the square of the velocity and it is defined as 
 \begin{equation}
F_\mathrm{aero,fs}(s) = \left(c_\mathrm{d,1} + c_\mathrm{d,2}\cdot \gamma(s) \right)\cdot v^{2}(s), \label{eq:faero}
\end{equation}
where $c_\mathrm{d,1}$ is the aerodynamic resistance coefficient in free stream, and the term $c_\mathrm{d,2}\cdot \gamma$ accounts for the influence of the sidewind caused by the path's curvature $\gamma(s)$ of the racing line. This effect is typical for open-wheel race cars \cite{ebbesen2017time} and the constant coefficient $c_\mathrm{d,2}$ quantifies its impact. In the next section, we model $F_\mathrm{aero,int}$.

\subsection{Drag Reduction Model}\label{dragRedModel}
We now introduce the model for the reduction of drag force $F_\mathrm{aero,int}$, which comes from the presence of another agent. Its equation reads
 \begin{equation}
F_\mathrm{aero,int}(s) = c_\mathrm{d,1}\cdot \delta(s) \cdot v^{2}(s), \label{eq:faeroint}
\end{equation}
where $\delta(s)$ is the reduction in percentage of the drag coefficient. The reduction exists because of the presence of the other agent, and therefore it also depends on its variables. In our model, it is the quantity which captures the interaction between them. For modeling purposes, we define the \textit{relative} gap time for agent $i$ as
\begin{equation}
t_{\mathrm{gap,rel},i}(s) = t_{i}(s) - t_{-i}(s) \qquad i\in\{A,B\},\label{tGapRel}
\end{equation}
where $-i$ means ``not the agent $i$'', such that we can describe the relative temporal position of each agent.

\begin{figure}
\centering
\begin{externalize}{cd_int_fit}
\begin{tikzpicture}[trim axis right]

\def\plotwidth{\columnwidth}%
\def\plotheight{4.8cm}%
\def\yshift{-0cm}%
\definecolor{mycolor1}{rgb}{0.00000,0.44700,0.74100}%
\definecolor{mycolor2}{rgb}{0.85000,0.32500,0.09800}%
\definecolor{mycolor3}{rgb}{0.92900,0.69400,0.12500}%

\begin{axis}[%
width=\plotwidth,
height=\plotheight,
at={(0,0)},
xmin=-0.25,
xmax=1.45,
xtick={-0.2, 0, 0.2, 0.4, 0.6, 0.8, 1, 1.2, 1.4},xlabel style={font=\color{white!15!black}},
xlabel={Relative gap time [s]},
 xlabel style={font=\color{white!15!black}},
xlabel={$t_\mathrm{gap,rel}$},
x unit= \unit{\second},
ymin=-0.05,
ymax=0.45,
ytick={0, 0.1, 0.2, 0.3, 0.4},
yticklabels={0, 10, 20, 30, 40},
ylabel style={font=\color{white!15!black}, 
yshift=\yshift},
ylabel={$\delta$},
y unit =  \unit{\percent},
axis background/.style={fill=white},
xmajorgrids,
ymajorgrids,
legend style={legend cell align=left, align=left, draw=white!15!black}
]

\addplot [color=gray, line width=0.5, mark=+, mark options={solid, gray, thick}] table[]{figures/cd_int/cd_int-2.tsv};
\addlegendentry{\cite{ravelli2021aerodynamics}}

\addplot [color=gray, line width=0.5, mark=x, mark options={solid, gray, thick}] table[]{figures/cd_int/cd_int-1.tsv};
\addlegendentry{\cite{guerrero2020aerodynamic}}

\addplot [color=gray, line width=1.8pt, loosely dashdotted] table[]{figures/cd_int/cd_int-4.tsv};
\addplot [color=gray, line width=1.8pt, loosely dashdotted, forget plot] table[]{figures/cd_int/cd_int-5.tsv};
\addlegendentry{Model}

\addplot [color=black, line width=0.7pt] table[]{figures/cd_int/cd_int-9.tsv};
\addplot [color=black, line width=0.7pt, forget plot] table[]{figures/cd_int/cd_int-10.tsv};
\addlegendentry{NN fit}
  
\end{axis}

\end{tikzpicture}%
\end{externalize}
\setlength{\abovecaptionskip}{\myfigskip}
\caption{Reduction in drag coefficient $\delta$. Comparison between literature data from \cite{ravelli2021aerodynamics} and \cite{guerrero2020aerodynamic}, model and \gls{nn} fit. On the $x$-axis we see the relative gap time.}
\label{fig:cd_int_fit}
\end{figure}
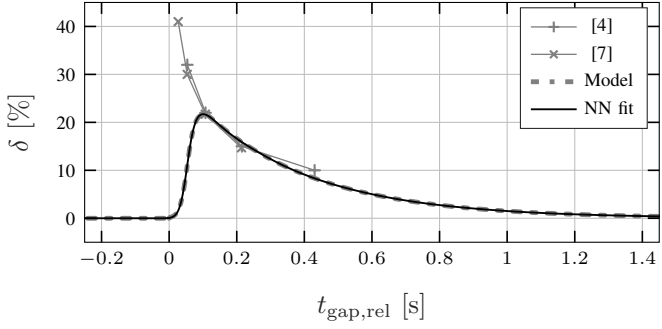

We characterize $\delta(s)$ based on the available literature and by employing modeling assumptions. Although the aerodynamic of single race cars is a widely studied topic, the investigations for a closely following vehicle in a wake flow are limited, and the lack of data is significant. However, a general trend for control purposes can be extrapolated, despite the differences in models (scaling, years) and conditions (different velocities, turbulence and Reynolds numbers). As reference for modern \gls{f1} cars, we consider the results of \cite{ravelli2021aerodynamics,guerrero2020aerodynamic}, where the authors make use of unscaled models of 2017 cars in \gls{cfd} simulations. Their data are summarized in \Cref{fig:cd_int_fit}. The following assumptions complete the model:
\begin{Ass} 
The perturbed flow behind the car changes instantaneously without any dynamic effect. 
\end{Ass}
\begin{Ass}  
The difference in velocity between the two cars is neglected, since in the literature only cars with the same velocity are considered.
\end{Ass} 
\begin{Ass}  
The vehicles are perfectly aligned and the effect on the drag reduction coming from the lateral shift is neglected.
\end{Ass} 
\begin{Ass} 
The reduction in drag coefficient scales with respect to the \textit{relative} gap time instead of the physical distance, and it is induced by the following reasoning. The drag coefficient is determined by the state of the perturbation in the flow, caused by the generated wake. Considering the same spacing at different velocities, the reduction in drag coefficient is not the same due to the different flow perturbations. For instance, at half of the velocity, the perturbation at the same distance is less pronounced, since the wake is shorter. On the other hand, using the same gap time at different velocities results in variable distances, compensating for the variable wake length. In space domain, we can interpret this as if the perturbation settles after a specific time for a selected point of the track. Moreover, the industrial partner confirmed the validity of this assumption without disclosing sensible data.
\end{Ass}
\begin{Ass}  
The drag reduction is lost as soon as the car begins the overtaking manoeuvre. When the car behind is close to overtaking, it will move to the side to avoid contact with the other car, exiting the wake. This means that it will experience free air although it is still behind, losing the drag reduction. The threshold in relative gap time for the beginning of an overtake is assumed to be $\qty{0.1}{\second}$.
\end{Ass} 

Based on the assumptions, we use elementary functions to create the model which captures the percentage of reduction in the drag coefficient as described in the literature. The result is shown in \Cref{fig:cd_int_fit}. The model is fitted with \acrfull{nn} techniques as in \cite{balerna2020time} with nonlinear activation functions. In this way we have a smooth and twice differentiable function, suited for nonlinear optimization techniques. It is described by the relation
 \begin{align}
 \delta_{i}(s) & = \mathcal{M} (t_{\mathrm{gap,rel},i}(s)) \nonumber \\
 &=  \mathcal{M} (t_{i}(s) - t_{-i}(s)), \label{eq:cd_int_NN}
 \end{align}
where $\mathcal{M}$ denotes the neural network function, and it is incorporated in both agents' models. With that particular shape, we can notice that if a car is far behind the other, $t_{\mathrm{gap,rel},i}$ is large and the drag reduction approaches zero. By diminishing the relative gap time, the car behind can decrease its drag, up to the threshold where the overtake begins. If the relative gap time becomes negative, i.e., the agent is in front, then no reduction in drag is present. For instance, if agent $B$ is $\qty{0.32}{\second}$ behind agent $A$ at $s^{\star}$, then
\begin{align*}
 t_{\mathrm{gap,rel},B}(s^{\star}) &= t_{B}(s^{\star})-t_{A}(s^{\star}) = \qty{0.32}{\second}, \\
 t_{\mathrm{gap,rel},A}(s^{\star}) &= t_{A}(s^{\star})-t_{B}(s^{\star}) = \qty{-0.32}{\second}, 
\end{align*}
and from \Cref{fig:cd_int_fit} we can see that at the point $s^{\star}$, $B$ can profit from a reduction of $\qty{11.6}{\percent}$ in the drag coefficient, whereas $A$ of $\qty{0}{\percent}$. 
In conclusion, we derived a model with the following properties: 
\begin{itemize}
\item It physically describes the effect of the wake on a following vehicle.
\item It is embeddable in an \gls{ocp} without the need of integer variables, thanks to its \gls{nn} fit. 
\item It is versatile, given the flexibility of the \gls{nn} fitting. E.g., it is easy to model cars with different aerodynamic properties.
\end{itemize}

\subsection{Statement of the Agents' Optimal Control Problems}
We now formulate the control problem for each agent. After presenting the detailed, continuous version, we state their discrete-space compact form. The latter is useful in \Cref{sec:gametheory} to explain the manipulations to which it is subjected to, in order to obtain the final problem formulation. For the moment, we consider a general cost function $J_{i}$ for each agent $i\in\{A,B\}$, which can, for instance, be a lap time minimization as in \cite{balerna2020time}.
\begin{problem}\label{continuousOCP}
The \gls{ocp} for the agent $i$ over one lap of an \gls{f1} race is
\begin{flalign*}
    \min_{P_\mathrm{k,i},P_\mathrm{f,i},P_\mathrm{brk,i}} J_{i}(s) 
\end{flalign*}
subject to the following constraints:
\begin{alignat*}{2}
		& \text{States:} \quad && \eqref{eq:time}, \eqref{eq:fuel}, \eqref{eq:battery}, \eqref{eq:velocity}, \\
		& \text{Power unit:} \quad && \eqref{eq:pkbound}, \eqref{eq:pfbound}, \eqref{eq:pbrkbound}, \eqref{eq:pengine}, \eqref{eq:propulsive}, \\
		& \text{External forces:} \quad && \eqref{eq:pext}, \eqref{eq:fext}, \eqref{eq:faero}, \eqref{eq:froll}, \eqref{eq:fslope}, \label{eq:faeroint}\\
		& \text{Boundary conditions:} \quad && \eqref{eq:tinit}, \eqref{eq:efbc}, \eqref{eq:ebbc}, \eqref{eq:ebbounds}, \eqref{eq:vbc},\\
		& \text{Coupling constraints:} \quad && \eqref{eq:cd_int_NN}. \\
	\end{alignat*}
\end{problem}
Note that the coupling constraints depend also on the state of the other agent $t_{-i}$. We transcribe the \gls{ocp} to obtain a finite-dimensional \gls{nlp} by employing the multiple shooting method \cite{rajasingham2021nonlinear} and the Euler forward integration scheme. The track is then discretized in $N$ steps denoted by $k$ as 
\begin{equation}
s\in[0,S] \quad \rightarrow \quad k \in \{1,\dots,N\},
\end{equation}
and to improve the readability, we introduce the following notation: The input and the state vectors for the step $k$ are 
\begin{align}
\mathbf{u}_{i}^{k} & =\begin{bmatrix}P_{\mathrm{k},i}^{k} & P_{\mathrm{f},i}^{k} & P_{\mathrm{brk},i}^{k}\end{bmatrix}, && k\in\{1,\dots,N-1\},\\
\mathbf{x}_{i}^{k} & =\begin{bmatrix} v_{i}^{k} & E_{\mathrm{f},i}^{k} & E_{\mathrm{b},i}^{k} & t_{i}^{k}\end{bmatrix}, && k\in\{1,\dots,N\},
\end{align}
and the vectors for the entire lap are 
\begin{align}
\mathbf{u}_{i} & =\begin{bmatrix} \mathbf{u}_{i}^{1}& \dots & \mathbf{u}_{i}^{N-1} \end{bmatrix}^{\intercal}\\
\mathbf{x}_{i} & =\begin{bmatrix} \mathbf{x}_{i}^{1} & \dots & \mathbf{x}_{i}^{N}\end{bmatrix}^{\intercal}.
\end{align}
\begin{problem}
The \glspl{nlp} versions of \Cref{continuousOCP} for the agents $ i\in\{A,B\}$ are
\begin{flalign*}
    \min_{\mathbf{x}_{i}, \mathbf{u}_{i}} J_{i}(\mathbf{x}_{i},\mathbf{u}_{i}, \mathbf{x}_{-i}) 
\end{flalign*}
\qquad \qquad subject to:
\begin{align*}
		\mathbf{g}_{i}(\mathbf{x}_{i},\mathbf{u}_{i}, \mathbf{x}_{-i})& \le 0, \\
		\mathbf{h}_{i}(\mathbf{x}_{i},\mathbf{u}_{i}, \mathbf{x}_{-i})& = 0.
\end{align*}
\end{problem}
The multiple shooting constraints are included in the vector of equality constraints $\mathbf{h}_{i}$. With this formulation, the dependency from the other agent becomes clear. The states $\mathbf{x}_{-i}$ are influencing the \gls{nlp} of agent $i$, but the optimization variables are only $\mathbf{x}_{i}$ and $\mathbf{u}_{i}$. For the investigation of the drag reduction, only the time vector of the other agent $t_{-i}^{k}$ is necessary. We stick to a more general form to show that with the proposed method, additional types of interaction are possible and easily implementable.

\subsection{Cost functions}\label{costFunctions}
So far, we considered a general cost function $J_{i}$ for each agent, which depends on the agent's states and inputs $\mathbf{x}_{i},\mathbf{u}_{i}$, as well as the other agent's states $\mathbf{x}_{-i}$. We stick to this notation also in \Cref{sec:gametheory}, to showcase the transferability of the method employed.

The multi-agent setting enables additional degrees of freedom regarding the choice of the cost functions. Depending on the racing situation, the strategic choices result in different objectives. However, it is essential to ensure that the problem is well-defined and to craft appropriate cost functions. Even in highly competitive scenarios, the main objective is still to drive as fast as possible, with a minor focus on strategic tasks, for instance not letting the other exploit the wake. Blocking or defending scenarios imply the inclusion of collision-free interactions, which are not the focus of this work. Hence, for the presented results, we rely on a lap time minimization of both agents, i.e., 
\begin{equation}
 J_{i}(s) =  t_{i}(S)  = \int_{0}^{S}\frac{1}{v_{i}(s)}\mathrm{d} s,
\end{equation}
with the discrete versions
\begin{equation}
 J_{i}(\mathbf{x}_{i}) =  t_{i}^{N}  = \sum_{k = 1}^{N}\frac{1}{v_{i}^{k}}.
\end{equation}

We highlight that the two agents behave egoistically, meaning that each one wants to minimize their own lap time, without actively cooperating or competing. Blocking or defensive maneuvers will not appear with the chosen cost function and the current interaction model. Although it might be counterintuitive, with the assumption of the drag reduction effect only, being behind is beneficial in terms of lap time. This means that an agent is not encouraged at blocking someone from overtaking. With a downforce interaction model together with a trajectory optimization, passive blocking maneuvers could be observed. In this case, being behind holds the disadvantage of losing downforce, forcing the agent behind to deviate from the optimal trajectory. Strategic behaviors can be included by augmenting the cost function on additional terms as in \cite{burger2022interaction} or by using the \gls{svo} model of \cite{schwarting2019social}. Last but not least, given the complexity of the system, a pure lap time minimization facilitates the analysis and the interpretation of the results.

\section{Game-theoretic Formulation}\label{sec:gametheory}
The problem description of \Cref{sec:modelinter} highlights the physical coupling and dependencies of the agents. Game theory arises as the natural choice to compute optimal strategies that explicitly take into account the interaction and the response of the other agent. In this section, we present the game-theoretic approach which allows to embed the interactions in the problem formulation. 

\subsection{Dynamic Stackelberg Game}\label{diffStackelberg}
The hierarchical structure of a Stackelberg game \cite{von2010market} is well-suited to model two \gls{f1} cars racing close to each other. In this type of game, one agent, the leader, makes the decision first, and the other agent, the follower, observes this decision and optimizes its own strategy accordingly. The leader is aware of the follower's response and takes it into account when taking its decision. 

In our problem setting, the car behind is the game leader. Intuitively, when fighting for a position, the agent behind attacks and the agent ahead has to defend himself. Even in our egoistic scenario, where pilot fights are not the focus, the agent behind still holds most of the decision power and thus more influence on the outcome of the game. Indeed, it can decide how to exploit the wake, whether to overtake and where. The agent ahead faces the consequences of these decisions. Without loss of generality, we chose agent $B$ to start behind, establishing it as the leader in the game-theoretic framework. This choice does not restrict alternative game formulations, e.g., where the race leader is also the leader of the game.  

Since our system is governed by differential equations, the problem can be formulated as a differential game \cite{bardi1999stochastic}, whose discretized version is a dynamic Stackelberg game \cite{bacsar1998dynamic}. The system dynamics are captured by 
\begin{align}
\mathbf{x}^{k+1}_{B} &= f_\mathrm{disc}(\mathbf{x}^{k}_{B}, \mathbf{u}^{k}_{B}, \mathbf{x}^{k}_{A}),\\
\mathbf{x}^{k+1}_{A} &= f_\mathrm{disc}(\mathbf{x}^{k}_{A}, \mathbf{u}^{k}_{A}, \mathbf{x}^{k}_{B}),
\end{align}
with $f_\mathrm{disc}$ the function of the discretized dynamics. 

The decisions of the agents in a classical Stackelberg game are of sequential nature. $B$ makes a decision knowing that $A$ will observe it and will respond optimally. This feedback relation is mathematically captured by a bilevel program, where the high-level program of the leader is constrained by the low-level program of the follower. Unlike simply solving two optimization problems in sequence, $B$ makes its decision with the awareness that $A$ will observe and respond to it.

\begin{problem}\label{prob:bilevel}
The bilevel program formulation of the dynamic Stackelberg game is
\begin{alignat*}{2}
 & \min_{\mathbf{x}_{B}, \mathbf{u}_{B}} J_{B}(\mathbf{x}_{B},\mathbf{u}_{B}, \mathbf{x}_{A}),\\
 & \text{subject to:} \\
		& \quad \quad \mathbf{g}_{B}(\mathbf{x}_{B},\mathbf{u}_{B}, \mathbf{x}_{A}) \le 0, \\
		& \quad \quad \mathbf{h}_{B}(\mathbf{x}_{B},\mathbf{u}_{B}, \mathbf{x}_{A}) = 0,\\
		& \quad \quad \{\mathbf{x}_{A},\mathbf{u}_{A}\} = \arg\min_{\mathbf{x}_{A}, \mathbf{u}_{A}}  J_{A}(\mathbf{x}_{A},\mathbf{u}_{A}, \mathbf{x}_{B}),\\
		& \quad \quad \text{subject to:} \\
		& \quad \quad \quad \quad \mathbf{g}_{A}(\mathbf{x}_{A},\mathbf{u}_{A}, \mathbf{x}_{B}) \le 0, \\
		& \quad \quad \quad \quad \mathbf{h}_{A}(\mathbf{x}_{A},\mathbf{u}_{A}, \mathbf{x}_{B}) = 0.
\end{alignat*}
\end{problem}
Since we designated $B$ as the leader, it corresponds to the high-level program in our bilevel program formulation.

\subsection{Reformulation as single-level \gls{nlp}}
An efficient solution method to solve bilevel programs is to replace the low-level program using its \gls{kkt} conditions \cite{pilecka2012combined}. This reduces the problem to a single-level \gls{nlp}, which can be solved using nonlinear solvers. This technique has proven to be effective to solve small- and medium-scale problems, mainly in \gls{mpc} applications \cite{burger2022interaction, schwarting2019social}. The drawbacks that this reformulation introduces are addressed in \Cref{sec:practical}.

The Lagrangian of the low-level program is defined as 
\begin{align}\label{lagrangian}
L(\mathbf{x}_{A},\mathbf{u}_{A}, \mathbf{x}_{B}) = & J_{A}(\mathbf{x}_{A},\mathbf{u}_{A}, \mathbf{x}_{B}) \nonumber\\
& +\boldsymbol{\lambda}^{\intercal}\cdot \mathbf{h}_{A}(\mathbf{x}_{A},\mathbf{u}_{A}, \mathbf{x}_{B})  \nonumber\\
& +\boldsymbol{\mu}^{\intercal}\cdot \mathbf{g}_{A}(\mathbf{x}_{A},\mathbf{u}_{A}, \mathbf{x}_{B}),
\end{align}
where $\boldsymbol{\lambda}$ and $\boldsymbol{\mu}$ are the vectors of Lagrange multipliers for the equality and inequality constraints, respectively. 
\begin{problem}\label{singleLevelProb}
The single-level \gls{nlp} reformulation for the bilevel program of \Cref{prob:bilevel} is
\begin{subequations}
\begin{alignat}{2}
& \min_{\mathbf{x}_{B}, \mathbf{u}_{B}, \mathbf{x}_{A}, \mathbf{u}_{A}, \boldsymbol{\lambda}, \boldsymbol{\mu}} J_{B}(\mathbf{x}_{B},\mathbf{u}_{B}, \mathbf{x}_{A}) + J_{A}(\mathbf{x}_{A},\mathbf{u}_{A}, \mathbf{x}_{B}), \nonumber\\
& \text{subject to:} \nonumber\\
		 &\quad \quad \mathbf{g}_{B}(\mathbf{x}_{B},\mathbf{u}_{B}, \mathbf{x}_{A})  \le 0, \nonumber\\
		 &\quad \quad \mathbf{h}_{B}(\mathbf{x}_{B},\mathbf{u}_{B}, \mathbf{x}_{A})  = 0,\nonumber\\
		 &\quad \quad \nabla_{\mathbf{x}_{A},\mathbf{u}_{A}}L(\mathbf{x}_{A},\mathbf{u}_{A}, \mathbf{x}_{B})  = 0, \label{eq:stationarity}\\
		 &\quad \quad \mathbf{g}_{A}(\mathbf{x}_{A},\mathbf{u}_{A}, \mathbf{x}_{B})  \le 0, \label{eq:primal1}\\
		 &\quad \quad \mathbf{h}_{A}(\mathbf{x}_{A},\mathbf{u}_{A}, \mathbf{x}_{B})  = 0, \label{eq:primal2}\\
		 &\quad \quad \boldsymbol{\mu} \ge 0, \label{eq:dual}\\
		 &\quad \quad \mu_{j}\cdot {g}_{A,j}(\mathbf{x}_{A},\mathbf{u}_{A}, \mathbf{x}_{B})   = 0, \quad j\in\{1,\dots,m\}. \label{eq:slackness}
\end{alignat}
\end{subequations}
\end{problem}
where $m$ is the number of inequality constraints of the inner problem, \eqref{eq:stationarity} the stationarity condition, \eqref{eq:primal1} and \eqref{eq:primal2} the primal feasibility, \eqref{eq:dual} the dual feasibility and \eqref{eq:slackness} the complementary slackness. \eqref{eq:stationarity} enforces to solve for stationary points of the low-level program, which can be local minima or maxima. In order to solve for a local minimum, we include the cost function of the low-level program in the high-level cost \cite{schwarting2019social}.

Summarizing, starting from a dynamic Stackelberg game in the form of a bilevel program, we recovered the structure of a single-level \gls{nlp}, maintaining the game-theoretic characteristics. 

\subsection{Practical Aspects of Implementation}\label{sec:practical}
The \gls{kkt} conditions used to replace the low-level program introduce computational challenges that have to be addressed. First, the complementary slackness of \eqref{eq:slackness} gives rise to a \gls{mpcc}. This type of mathematical programs are generally difficult to solve, since they violate constraint qualifications (such as \gls{licq} and \gls{mfcq}) at feasible points. Using the Scholtes' relaxation scheme \cite{scholtes2001convergence} for the complementary slackness 
\begin{equation}
\mu_{j}\cdot {g}_{A,j}(\mathbf{x}_{A},\mathbf{u}_{A}, \mathbf{x}_{B})  \ge -\varepsilon, \quad j\in\{1,\dots,m\},
\end{equation}
with $\varepsilon\ge 0$, the problem is no more a \gls{mpcc}. This relaxation scheme recovers and ensures \gls{mfcq} at feasible points \cite{hoheisel2013theoretical}, and with a careful choice of the parameter $\varepsilon$, it can be easily tackled by off-the-shelf solvers.

Second, the computational time and the local minima are additional challenges to overcome. It is widely known that the use of warm starts for \glspl{nlp} is beneficial. For this reason, we compute the corresponding free-stream solution for each agent, whose use is threefold:
\begin{itemize}
\item As a measure to avoid the convergence towards highly suboptimal local minima, which are nevertheless computationally accessible. Moreover, given the nature of the drag reduction model, local minima with a strong attraction region are to be expected. For instance, an agent which remains in the wake without overtaking.
\item To speed up the solving time.
\item As a benchmark for the single agent solutions in the game (more on this in \Cref{sec:resultsDef}).
\end{itemize}
The game with two agents of \Cref{singleLevelProb} features more than $\num{17000}$ variables and $\num{22800}$ constraints. Despite its considerable size, the computational time for a single game ranges from $\qty{0.95}{\second}$ to $\qty{3}{\minute}$ on a laptop. The problem is parsed with CasADi \cite{andersson2012casadi}, whose algorithmic differentiation properties are used to compute the gradient of \eqref{lagrangian}, and then solved with IPOPT \cite{wachter2006implementation}.

\section{Results}\label{sec:results}
In this section, we showcase the potential of the developed optimization framework through the application on some case studies. First, we analyze the outcomes of a single game which exhibits an overtake, comparing the policy with the state-of-the-art single-agent optimal solution. Second, we study how the agents' energy budgets affect the lap time for different boundary conditions. Finally, we analyze the sensitivity of the lap time improvement towards initial gap time and allocated energy. Some plots are normalized for confidentiality reasons. 

\subsection{Definitions}\label{sec:resultsDef}
Before delving into the results, it is necessary to provide some explanations to ensure a proper interpretation. Results are shown for the two agents $A$ and $B$, and to represent their trajectories we use red and blue, respectively.

We use the free-stream solutions to benchmark the trajectories resulting from the games. Since the free-stream cases receive the same boundary conditions, they correspond to the optimal single-agent case, where the interaction does not exist. Additionally, considering only the drag reduction, the free-stream solutions also represent the worst-case scenario for the chosen cost functions $J_{i}(\mathbf{x}_{i}) =  t_{i}^{N}$. Indeed, if even for only one step the agent can profit from a reduction, its lap time must be lower than its free-stream case. We denote the lap time of the free-stream solutions as $t_{\mathrm{fs},i}$, the lap times resulting from the game as $t_{\mathrm{g},i}$, and the improvement 
\begin{equation}
\Delta t_{\mathrm{lap},i} = t_{\mathrm{g},i} - t_{\mathrm{fs},i}. \label{eq:improvementLap}
\end{equation}
When speaking of gap time, we use the \textit{relative} gap time of agent $B$ as a reference:
\begin{equation}
 t_\mathrm{gap}^{k} = t_{\mathrm{gap,rel},B}^{k} \label{eq:gapTimePlots}
\end{equation}
This means that if $t_\mathrm{gap}^{k}\ge 0$, the agent $B$ is behind agent $A$, and vice versa.

The initial temporal position of the agents is defined by the initial gap time. In our problem setting, agent $B$ starts behind. This means that its initial time is greater than the one of $A$: 
\begin{equation}
t_\mathrm{init,B} \ge t_\mathrm{init,A}.
\end{equation}
Since the nature of the drag interaction is given by the gap time and not by the absolute time, we can set $t_\mathrm{init,A} = \qty{0}{\second}$ without loss of generality. The boundary conditions are fully defined by specifying the \textit{initial} gap time
\begin{equation}
 t_\mathrm{gap,init} = t_\mathrm{init,B} - t_\mathrm{init,A}.
\end{equation}

Summarizing, the boundary conditions that define a game are: the initial velocity of each agent $v_{\mathrm{init},i}$, the battery initial energy content $E_{\mathrm{b,init},i}$, the battery allocated energy $\Delta E_{\mathrm{b},i}$, the fuel target $E_{\mathrm{f,target},i}$ and the initial gap time $t_\mathrm{gap,init}$. For the presented case studies, carried out on the Bahrain International Circuit, we always consider $v_{\mathrm{init},i} = \qty{300}{\kilo\meter\per\hour}$, $E_{\mathrm{b,init},i} = \qty{2}{\mega\joule}$, and equal fuel targets $E_{\mathrm{f,target},i}$. The remaining boundary conditions that we vary among the case studies are condensed in the vector
\begin{equation}
\BCvector = \begin{bmatrix} \Delta E_{\mathrm{b},A} & \Delta E_{\mathrm{b},B} & t_\mathrm{gap,init}\end{bmatrix}^{\intercal}.
\end{equation}	

\subsection{Energy Management Strategy: Overtaking Scenario}\label{S:casestudy1}
The single game analyzed here is obtained with the boundary conditions
\begin{equation}\label{BC1}
\BCvector = \begin{bmatrix} \qty{0}{\mega\joule} &  \qty{-1.4}{\mega\joule} & \qty{0.2}{\second}\end{bmatrix}^{\intercal},
\end{equation}
meaning that $A$ is required to use a charge-sustaining strategy, $B$ is allowed to discharge \qty{1.4}{\mega\joule} of battery energy, and $B$ passes the start line \qty{0.2}{\second} after $A$. In \Cref{fig:singleGameSol}, we can see the velocity trajectories of both agents, their aerodynamic drag powers $P_{\mathrm{aero},i}$ and the gap time as defined in \eqref{eq:gapTimePlots}. The drag power plots show two lines: the total drag power coming from $F_\mathrm{aero,tot}$, in the agent's color, and the drag power resulting from $F_\mathrm{aero,fs}$, in gray. Note that the latter does not correspond to the free-stream solution trajectory, but it is the drag force of \eqref{eq:faero}, as if there were no vehicle ahead.
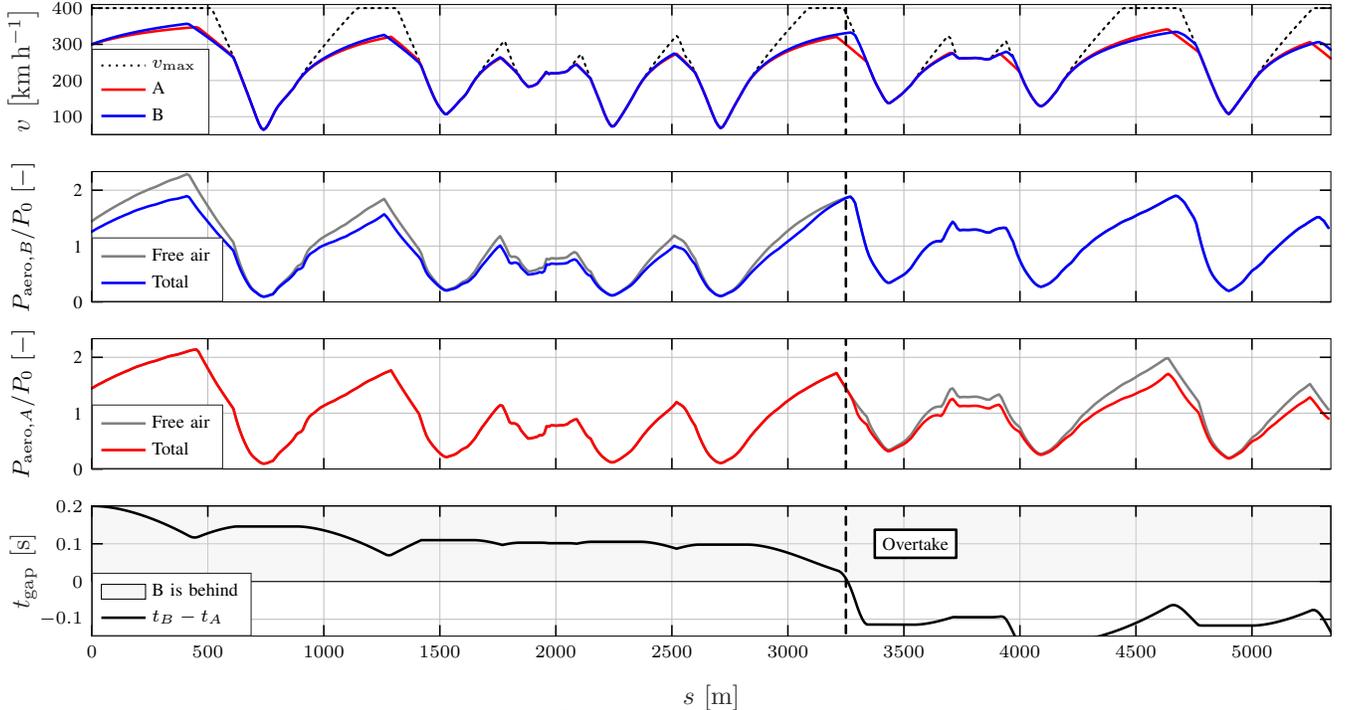
\begin{figure*}
\centering
\begin{externalize}{singleGameSol}
\definecolor{mycolor1}{rgb}{1.00000,0.00000,1.00000}%

\begin{tikzpicture}[trim axis right]

\def\plotwidth{2\columnwidth}%
\def\plotheight{3.4cm}%
\def\plotoffset{(\plotheight - 1cm)}%
\def\yshift{0.3cm}%
\def\yshiftT{-0.15cm}%

\begin{axis}[%
width=\plotwidth,
height=\plotheight,
at={(0,0)},
xmin=0,
xmax=5340,
xtick={0,500,1000,1500,2000,2500,3000,3500,4000,4500,5000},
xticklabels={{}},
ymin=50,
ymax=410,
ylabel style={font=\color{white!15!black},yshift=0, font=\small},
ylabel={$v$},
y unit= \unit{\kilo\meter\per\hour},
axis background/.style={fill=white},
xmajorgrids,
ymajorgrids,
legend style={at={(0,0)}, anchor=south west, legend cell align=left, align=left, draw=white!15!black}
]
\addplot [color=black, dotted, line width=0.8pt]
  table[]{figures/CS1/CS1_1-1.tsv};
\addlegendentry{$v_\mathrm{max}$}

\addplot [color=red, line width=1.0pt]
  table[]{figures/CS1/CS1_1-2.tsv};
\addlegendentry{A}

\addplot [color=blue, line width=1.0pt]
  table[]{figures/CS1/CS1_1-3.tsv};
\addlegendentry{B}

\draw[dashed, black, line width=1.0pt] (axis cs:3250,0) -- (axis cs:3250,400);

\end{axis}

\begin{axis}[%
width=\plotwidth,
height=\plotheight,
at={(0,-\plotoffset)},
xmin=0,
xmax=5340,
xticklabels={{}},
xlabel style={font=\color{white!15!black}},
ymin=-0.0329906968132237,
ymax=700,
ytick={0, 300, 600},
yticklabels={0, 1, 2},
ylabel style={font=\color{white!15!black},yshift=\yshift, font=\small},
ylabel={$P_{\mathrm{aero},B}/P_{0}$},
y unit= -,
axis background/.style={fill=white},
title style={font=\bfseries},
xmajorgrids,
ymajorgrids,
legend style={at={(0,0)}, anchor=south west, legend cell align=left, align=left, draw=white!15!black}
]

\addplot [color=gray, line width=1.0pt, fill opacity=0.3]
  table[]{figures/CS1/CS1_1-5.tsv};
\addlegendentry{Free air}

\addplot [color=blue, line width=1.0pt]
  table[]{figures/CS1/CS1_1-4.tsv};
\addlegendentry{Total}

\draw[dashed, black, line width=1.0pt] (axis cs:3250,0) -- (axis cs:3250,756);

\end{axis}

\begin{axis}[%
width=\plotwidth,
height=\plotheight,
at={(0,-2*\plotoffset)},
xmin=0,
xmax=5340,
xticklabels={{}},
xlabel style={font=\color{white!15!black}},
ytick={0, 300, 600},
yticklabels={0, 1, 2},
ymin=-0.0440630017722931,
ymax=700,
ylabel style={font=\color{white!15!black},yshift=\yshift, font=\small},
ylabel={$P_{\mathrm{aero},A}/P_{0}$},
y unit= -,
axis background/.style={fill=white},
title style={font=\bfseries},
xmajorgrids,
ymajorgrids,
legend style={at={(0,0)}, anchor=south west, legend cell align=left, align=left, draw=white!15!black}
]

\addplot [color=gray, line width=1.0pt, fill opacity=0.3]
  table[]{figures/CS1/CS1_1-8.tsv};
\addlegendentry{Free air}

\addplot [color=red, line width=1.0pt]
  table[]{figures/CS1/CS1_1-7.tsv};
\addlegendentry{Total}

\draw[dashed, black, line width=1.0pt] (axis cs:3250,0) -- (axis cs:3250,756);

\end{axis}

\begin{axis}[%
width=\plotwidth,
height=\plotheight,
at={(0,-3*\plotoffset)},
xmin=0,
xmax=5340,
xtick={0,500,1000,1500,2000,2500,3000,3500,4000,4500,5000},
xlabel style={font=\color{white!15!black}},
xlabel={$s$},
x unit = \unit{\meter},
ymin=-0.18,
ymax=0.2,
ylabel style={font=\color{white!15!black}, yshift=\yshiftT, font=\small},
ylabel={$t_\mathrm{gap}$},
y unit = \unit{\second},
axis background/.style={fill=white},
xmajorgrids,
ymajorgrids,
legend style={at={(0,0)}, anchor=south west, legend cell align=left, align=left, draw=white!15!black}
]

\addplot[area legend, draw=black, fill=white!90!black, fill opacity=0.3]
table[] {figures/CS1/CS1_1-10.tsv}--cycle;
\addlegendentry{B is behind}

\addplot [color=black, line width=1.0pt]
  table[]{figures/CS1/CS1_1-11.tsv};
\addlegendentry{$t_{B} - t_{A}$}

\draw[dashed, black, line width=1.0pt] (axis cs:3250,-1) -- (axis cs:3250,1);
\node[draw=black, line width = 1.0pt, fill=white, inner sep=3pt, text=black, font=\scriptsize] (boundingbox) at (axis cs:3550,0.1) {Overtake};

\end{axis}

\end{tikzpicture}%
\end{externalize}
\setlength{\abovecaptionskip}{\myfigskip}
\caption{Trajectories resulting from the solution of a single game. From top to bottom: Velocity trajectories for both agents, aerodynamic drag power of agent $B$, aerodynamic drag power of agent $A$, and gap time. The dashed vertical line represents the overtake location. For confidentiality reasons, the actual drag powers are normalized.}
\label{fig:singleGameSol}
\end{figure*}
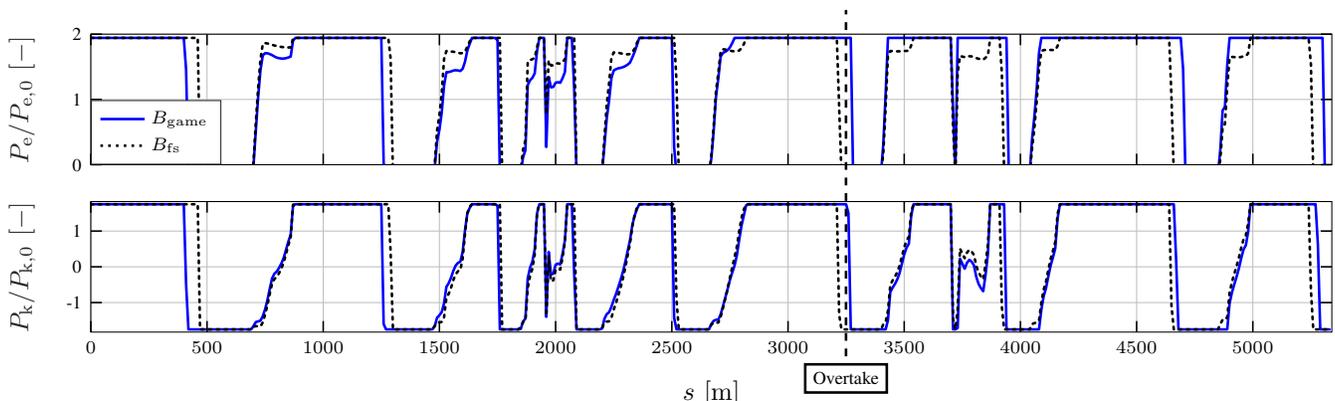
\begin{figure*}
\centering
\begin{externalize}{singleGameSol_compareFS}
\begin{tikzpicture}[trim axis right]

\def\plotwidth{2\columnwidth}%
\def\plotheight{3.4cm}%
\def\plotoffset{(\plotheight - 1cm)}%
\def\yshift{0.3cm}%

\begin{axis}[%
width=\plotwidth,
height=\plotheight,
at={(0,0)},
xmin=0,
xmax=5340,
xtick={0,500,1000,1500,2000,2500,3000,3500,4000,4500,5000},
xticklabels={{}},
ymin=0,
ymax=400,
ytick={0, 200, 400},
yticklabels={0, 1, 2},
ylabel style={font=\color{white!15!black},yshift=\yshift, font=\small},
ylabel={$P_\mathrm{e}/P_{\mathrm{e},0}$},
y unit = -,
axis background/.style={fill=white},
xmajorgrids,
ymajorgrids,
legend style={at={(0,0)}, anchor=south west, legend cell align=left, align=left, draw=white!15!black}
]
\addplot [color=blue, line width=1.0pt]
  table[]{figures/CS1/CS1_2-1.tsv};
\addlegendentry{$B_\mathrm{game}$}

\addplot [color=black, line width=1.0pt, dotted]
  table[]{figures/CS1/CS1_2-2.tsv};
\addlegendentry{$B_\mathrm{fs}$}

\end{axis}

\begin{axis}[%
width=\plotwidth,
height=\plotheight,
at={(0,-\plotoffset)},
xmin=0,
xmax=5340,
xtick={0,500,1000,1500,2000,2500,3000,3500,4000,4500,5000},
xlabel={$s$},
x unit = \unit{\meter},
ymin=-365,
ymax=365,
ytick={-200, 0, 200},
yticklabels={-1, 0, 1},
ylabel style={font=\color{white!15!black},yshift=\yshift, font=\small},
ylabel={$P_\mathrm{k}/P_{\mathrm{k},0}$},
y unit = -,
axis background/.style={fill=white},
xmajorgrids,
ymajorgrids,
]
\addplot [color=blue, line width=1.0pt]
  table[]{figures/CS1/CS1_2-3.tsv};

\addplot [color=black, line width=1.0pt, dotted]
  table[]{figures/CS1/CS1_2-4.tsv};

\end{axis}

\draw[dashed, black, line width=1.0pt] (9.9,-3.3) -- (9.9,1.9);
\node[draw=black, line width = 1.0pt, fill=white, inner sep=3pt, text=black, font=\scriptsize] (boundingbox) at (10.6,-3.2) {Overtake};

\end{tikzpicture}%
\end{externalize}
\setlength{\abovecaptionskip}{\myfigskip}
\caption{Engine and \gls{mguk} powers trajectories of agent $B$. Here we compare the result of the game with its corresponding free-stream solution without interactions. The dashed vertical line represents the overtake location. For confidentiality reasons, the engine and \gls{mguk} powers are normalized.}
\label{fig:singleGameSol_compareFS}
\end{figure*}

We first assess the physical validity of the drag reduction model and its impact on the framework. At $\qty{3250}{\meter}$ the gap time becomes negative, meaning that $B$ performed an overtake. In the drag powers plots, we observe that the model of drag reduction works as expected. As long as agent $A$ is ahead, its total drag power corresponds exactly to the drag in free air. After $A$ is overtaken, it begins to experience a reduced drag from the interaction with $B$. Exactly the opposite is valid for agent $B$. Additionally, the model demonstrates that overtaking manoeuvers occur towards the end of a straight, where the velocity difference is maximized. This is analogous to real \gls{f1} racing, when pilots exploit this interaction to reduce the gap or to overtake.

The impact on the energy management is analyzed in \Cref{fig:singleGameSol_compareFS}, where we compare the engine and \gls{mguk} power trajectories of $B$ with its free-stream solution $B_\mathrm{fs}$. This represents a direct benchmark with the single-agent optimal solution. We point out that both solutions received the same energy targets. In the multi-agent scenario, $B$ saves $\qty{1.73}{\mega\joule}$ of energy from the drag reduction. Given the energy surplus, one would expect that the saved energy is evenly redistributed over the whole lap, delaying the power cuts at the end of every straight. Interestingly, we notice a different strategy when another agent is present. In fact, before $B$ overtakes, its power cuts occur earlier than the ones of the free-stream solution. Additionally, $B$ can sustain higher peak velocities thanks to the reduction in drag, although in this section its energy consumption is lower than that of $A$ (not shown here). After the overtake, the energy management trend inverts: The power cuts of $B$ take place later than the ones of $B_\mathrm{fs}$, exploiting the previously saved energy.

\begin{table}
\begin{center}
\caption{Comparison of lap times in the game\\with free-stream solutions.}
\label{tab1}
\begin{tabular}{l  r  r}
\toprule
\textbf{Lap time} & \multicolumn{1}{c}{\textbf{A}} & \multicolumn{1}{c}{\textbf{B}}\\
\midrule
Free stream $t_\mathrm{fs}$ & $\qty{91.917}{\second}$ & $\qty{91.547}{\second}$\\
Game $t_\mathrm{g}$ & $\qty{91.518}{\second}$ & $\qty{91.172}{\second}$\\
Improvement $\Delta t_{\mathrm{lap}}$ & $\qty{-0.399}{\second}$ & $\qty{-0.375}{\second}$\\ 
\bottomrule
\end{tabular}
\end{center}
\end{table}
\Cref{tab1} compares the lap times of the free-stream solutions of $A$ and $B$ with the lap times achieved in the game. Since the agents received different energy targets, their free-stream lap times are also different. Thanks to the drag reduction, $A$ and $B$ are each able to reduce their lap times by $\qty{0.399}{\second}$ and $\qty{0.375}{\second}$, respectively. $A$ improved its lap time more than $B$, although the latter could profit from the wake for a longer section. We have to keep in mind that they are compared to the respective free-stream cases, and a direct comparison of the lap time gain is possible only for the same boundary conditions. Moreover, the combination of the agents' allocated energy budgets has also an impact on the lap time gain, and this effect is discussed in \Cref{sec:allocation}. 

\begin{table}
\begin{center}
\caption{Lap time improvements for circuits with\\different characteristics.}
\label{tab2}
\begin{tabular}{l  r  r r}
\toprule
 & \multicolumn{1}{c}{\textbf{Hungary}} & \multicolumn{1}{c}{\textbf{Bahrain}} & \multicolumn{1}{c}{\textbf{Monza}}\\
\midrule
$\Delta t_{\mathrm{lap},A}$  & $\qty{-0.123}{\second}$ & $\qty{-0.399}{\second}$ & $\qty{-0.528}{\second}$\\
$\Delta t_{\mathrm{lap},B}$  & $\qty{-0.228}{\second}$ & $\qty{-0.375}{\second}$ & $\qty{-0.692}{\second}$\\
\midrule
Length & $\qty{4381}{\meter}$ & $\qty{5412}{\meter}$ & $\qty{5793}{\meter}$ \\
\bottomrule
\end{tabular}
\end{center}
\end{table}
\gls{f1} circuits have different characteristics. The Hungaroring features shorter straights than Bahrain, whereas the Autodromo Internazionale di Monza is known for its long straights, where pilots reach record peak velocities. \Cref{tab2} shows the lap time improvement across these circuits for the same boundary conditions. In each case there is an overtake, as both agents can improve the lap time. Compared to Bahrain, we can notice that the lap time improvements are smaller for the Hungaroring but greater in Monza, accordingly to their features. We are aware that different circuit lengths affects the time to exploit the drag reduction, enhancing or reducing the potential lap time gain. While this holds true for the shorter Hungaroring, we notice that in Monza higher lap time gains are achieved, despite having a comparable length with Bahrain. This shows that our framework takes into account the individual track characteristics.

These gains hold significant practical importance in \gls{f1}, where race results are often decided by fractions of a second. The achieved lap time improvements are effective only if the energy is optimally managed, with evident differences from the expected single-agent strategy. This showcases the importance of including the energy management when considering the interaction between agents. 

\subsection{Energy Allocation to Improve the Lap Time}\label{sec:allocation}
In this section, the focus lies on how the allocated energy affects the lap time improvement of both agents. To this end, we first fix the energy budget for both agents and analyze how the lap time can be improved for agent $A$, varying the initial gap time. Thereafter, the impact of the energy allocation of $B$ is discussed, for two different energy targets of $A$. 

\Cref{fig:lapTimeVarA} shows the lap time improvement $\Delta t_{\mathrm{lap},A}$ for initial gap times ranging from $\qty{0}{\second}$ to $\qty{1.4}{\second}$. The considered boundary condition vector reads
\begin{equation}
\BCvector = \begin{bmatrix} \qty{0}{\mega\joule} &  \qty{-1.8}{\mega\joule} & \tilde t_\mathrm{gap,init}\end{bmatrix}^{\intercal},
\end{equation}
where $\tilde t_\mathrm{gap,init} \in [\qty{0}{\second},\qty{1.4}{\second}]$. Larger initial gap times indicate that $A$ starts with added advantage. By inspecting the single solutions, we can distinguish the region where $A$ is overtaken by $B$ from where it is not, highlighted in gray.

For $t_\mathrm{gap,init}$ greater than \qty{0.8}{\second}, $B$ does not overtake, although it has allocated $\qty{1.8}{\mega\joule}$ of energy more than $A$. Consequently, this leads to a scenario where A cannot improve its lap time as it cannot benefit anywhere from a reduction in drag. Indeed, the curve results flat within the region where no overtake occurs. An exception takes place at $t_\mathrm{gap,init} = \qty{0.9}{\second}$, where we observe a slight increase in the lap time of $A$. This is given by numerical issues, for instance the relaxation of the complementarity constraints or a local minimum.

Conversely, where $B$ overtakes, we observe a change in the trend. Given these fixed energy budgets but only reducing the initial gap time, $B$ successfully overtakes $A$. The lap time improvement $\Delta t_{\mathrm{lap},A}$ varies, based on the overtake location, as indicated in the figure. The earlier the overtake occurs, the more lap time $A$ can gain, because it spends more time in $B$'s wake increasing the amount of saved energy. A deviation from the trend is observed for initial gap times of $t_\mathrm{gap,init} = \qty{0.4}{\second}$ to $t_\mathrm{gap,init} = \qty{0.8}{\second}$. It seems that although $A$ is overtaken, it does not improve its lap time. However, these are solutions where $B$ overtakes at the last discrete step, making the lap time improvement not noticeable. 
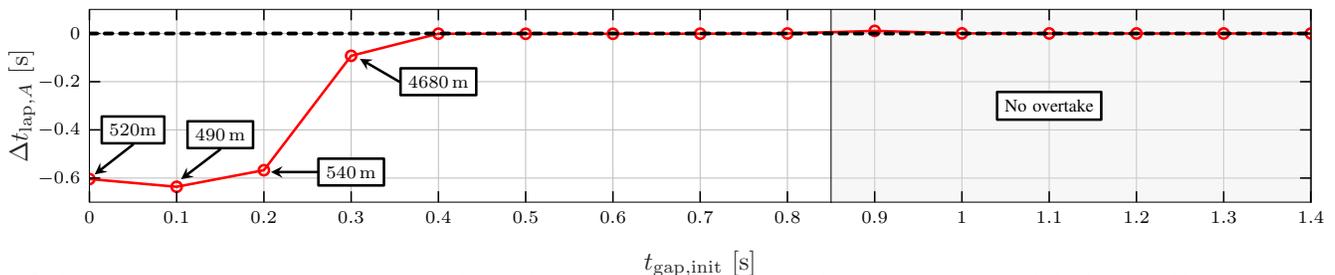
\begin{figure*}
\centering
\begin{externalize}{lapTimeVarA}
\begin{tikzpicture}[trim axis right]

\def\plotwidth{2\columnwidth}%
\def\plotheight{4.3cm}%
\def\plotoffset{(\plotheight + 0.4cm)}%
\def\yshift{-0cm}%

\begin{axis}[%
width=\plotwidth,
height=\plotheight,
at={(0,0)},
xmin=0,
xmax=1.4,
xlabel style={font=\color{white!15!black}},
xlabel={$t_\mathrm{gap,init}$},
x unit = \unit{\second},
ymin=-0.7,
ymax=0.1,
ylabel style={font=\color{white!15!black},yshift=\yshift},
ylabel={$\Delta t_{\mathrm{lap},A}$},
y unit = \unit{\second},
axis background/.style={fill=white},
xmajorgrids,
ymajorgrids,
]

\addplot[area legend, draw=black, fill=white!90!black, fill opacity=0.3]
table[] {figures/CS2/CS2_1-3.tsv}--cycle;

\addplot [color=red, line width=1.0pt, mark=o, mark options={solid, red}]
  table[]{figures/CS2/CS2_1-1.tsv};

\addplot [color=black, dashed, line width=1.5pt, forget plot]
  table[]{figures/CS2/CS2_1-2.tsv};

\node[draw=black, line width = 1.0pt, fill=white, inner sep=3pt, font=\scriptsize] (boundingbox1) at (axis cs:1.1,-0.3) {No overtake};
\node[draw=black, line width = 1.0pt, fill=white, inner sep=3pt, font=\scriptsize] (boundingbox2) at (axis cs:0.4,-0.2) {\qty{4680}{\meter}};
\node[draw=black, line width = 1.0pt, fill=white, inner sep=3pt, font=\scriptsize] (boundingbox3) at (axis cs:0.3,-0.575) {\qty{540}{\meter}};
\node[draw=black, line width = 1.0pt, fill=white, inner sep=3pt, font=\scriptsize] (boundingbox4) at (axis cs:0.15,-0.42) {\qty{490}{\meter}};
\node[draw=black, line width = 1.0pt, fill=white, inner sep=3pt, font=\scriptsize] (boundingbox5) at (axis cs:0.05,-0.4) {$520\unit{m}$};

\draw[->, line width = 1.0pt, >=stealth] (boundingbox2.west) -- (axis cs:0.31,-0.105);
\draw[->, line width = 1.0pt, >=stealth] (boundingbox3.west) -- (axis cs:0.208,-0.575);
\draw[->, line width = 1.0pt, >=stealth] (boundingbox4.south) -- (axis cs:0.105,-0.605);
\draw[->, line width = 1.0pt, >=stealth] (boundingbox5.south) -- (axis cs:0.006,-0.585);

\end{axis}

\end{tikzpicture}%
\end{externalize}
\setlength{\abovecaptionskip}{\myfigskip}
\caption{Lap time improvement of $A$ as a function of the initial gap time. The region where $B$ does not overtake is highlighted in gray. The tags indicate the overtake locations of $B$.}
\label{fig:lapTimeVarA}
\end{figure*}
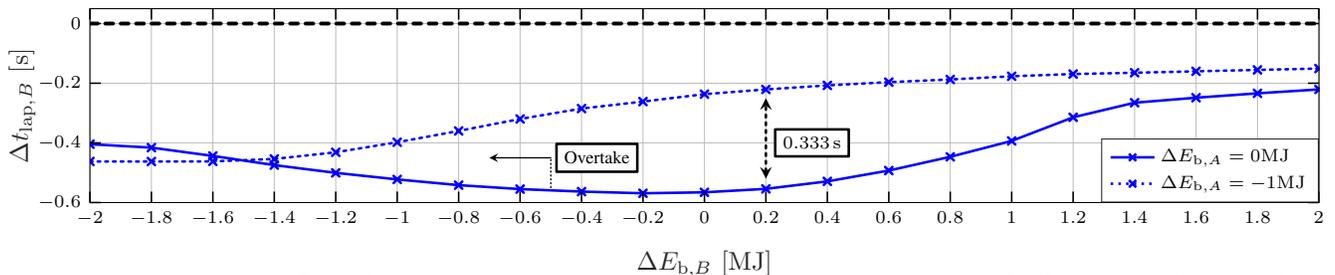
\begin{figure*}
\centering
\begin{externalize}{lapTimeVarB}
\begin{tikzpicture}[trim axis right]

\def\plotwidth{2\columnwidth}%
\def\plotheight{4.3cm}%
\def\plotoffset{(\plotheight + 0.4cm)}%
\def\yshift{-0cm}%

\begin{axis}[%
width=\plotwidth,
height=\plotheight,
at={(0,0)},
xmin=-2,
xmax=2,
xlabel style={font=\color{white!15!black}},
xlabel={$\Delta E_{\mathrm{b},B}$},
x unit =\unit{\mega\joule},
ymin=-0.6,
ymax=0.05,
ylabel style={font=\color{white!15!black},yshift=\yshift},
ylabel={$\Delta t_{\mathrm{lap},B}$},
y unit = \unit{\second},
axis background/.style={fill=white},
xmajorgrids,
ymajorgrids,
legend style={at={(1,0)}, anchor=south east, legend cell align=left, align=left, draw=white!15!black}
]

\addplot [color=blue, line width=1.0pt, mark=x, mark options={solid, blue}]
  table[]{figures/CS2/CS2_2-2.tsv};
\addlegendentry{$\Delta E_{\mathrm{b},A} = 0\unit{MJ}$}

\addplot [color=blue, dotted, line width=1.0pt, mark=x, mark options={solid, blue}]
  table[]{figures/CS2/CS2_2-1.tsv};
\addlegendentry{$\Delta E_{\mathrm{b},A} = -1\unit{MJ}$}

\addplot [color=black, dashed, line width=1.5pt, forget plot]
  table[]{figures/CS2/CS2_2-3.tsv};
  
\draw[black, line width=0.5pt, densely dotted] (axis cs:-0.5,-0.55) -- (axis cs:-0.5,-0.45);   
\draw[->, solid, black, line width=0.5pt, >=stealth] (axis cs:-0.5,-0.45) -- (axis cs:-0.7,-0.45);   
\node[draw=black, line width = 1.0pt, fill=white, inner sep=3pt, font=\scriptsize] (boundingbox1) at (axis cs:-0.35,-0.45) {Overtake};

\draw[<->, dotted, black, line width=1pt, >=stealth] (axis cs:0.2,-0.52) -- (axis cs:0.2,-0.25);   
\node[draw=black, line width = 1.0pt, fill=white, inner sep=3pt, font=\scriptsize] at (axis cs:0.35,-0.4) {\qty{0.333}{\second}};

\end{axis}

\end{tikzpicture}%
\end{externalize}
\setlength{\abovecaptionskip}{\myfigskip}
\caption{Lap time improvement of $B$ as a function of its allocated energy $\Delta E_{\mathrm{b},B}$. The two curves are the result of different energy allocations of $A$. The indicated overtake region belongs to the $\Delta E_{\mathrm{b},A} = \qty{0}{\mega\joule}$ case, since for the other no overtake is taking place.}
\label{fig:lapTimeVarB}
\end{figure*}

To further demonstrate the significant impact of the energy allocation, we consider now the results in \Cref{fig:lapTimeVarB}, obtained with the boundary conditions
\begin{equation}
\BCvector = \begin{bmatrix} \Delta \tilde E_{\mathrm{b},A} & \Delta\tilde E_{\mathrm{b},B} & \qty{0.6}{\second}\end{bmatrix}^{\intercal},
\end{equation}
where $\Delta \tilde E_{\mathrm{b},A}\in \{\qty{-1}{\mega\joule},\qty{0}{\mega\joule}\}$ and $\Delta\tilde E_{\mathrm{b},B}\in[\qty{-2}{\mega\joule},\qty{2}{\mega\joule}]$. Moving to the left on the plot means that $B$ can use more battery energy for the current lap.

We first analyze the curve for $\Delta  E_{\mathrm{b},A}=\qty{0}{\mega\joule}$. $B$ overtakes only for this case, and the corresponding region is indicated in the figure. Starting where $B$ overtakes at $\Delta E_{\mathrm{b},B} = \qty{-0.6}{\mega\joule}$, and moving to the left (thus increasing the allocated battery energy), we observe a decrease in the lap time gain. Although counterintuitive, this is motivated by the following fact. With more energy, $B$ is faster and has to overtake earlier and earlier to be lap time optimal. The earlier the overtake occurs, the less time is spent in the wake of $A$, reducing the saved drag energy and mitigating the potential gain.
In the region where $B$ does not overtake, the gain in lap time reaches a maximum at $\Delta E_{\mathrm{b},B} = \qty{-0.2}{\mega\joule}$ with $\qty{0.569}{\second}$. By reducing the energy that $B$ can use (moving to the right), we observe that the gain in lap time decreases again. $B$ is slower with less energy, increasing the gap time from $A$. As a consequence, it cannot sufficiently exploit the wake effect, reducing the potential lap time gain.

The curve for which $A$ allocates $\Delta  E_{\mathrm{b},A} = \qty{-1}{\mega\joule}$ shows a similar trend. Compared to the previous case, we observe the evident shift towards higher energy budgets for $B$. Since $A$ uses more energy, it achieves lower lap times. Therefore, $B$ must also invest more energy to keep up and exploit the wake effect. Moreover, $B$ does not overtake at all in this scenario, and indeed, moving to the left we do not observe the same decrease in lap time gain previously described. For this curve, the maximum lap time gain of $\qty{0.462}{\second}$ is reached at $\Delta E_{\mathrm{b},B} = \qty{-2}{\mega\joule}$.

Robust strategies can be identified based on the combinations of energy budgets between agents. With the underlying assumption that they optimally manage the energy, we can make the following considerations: Even though the choice $\Delta E_{\mathrm{b},B} = \qty{0.2}{\mega\joule}$ shows one of the largest lap time gains, it is suboptimal in terms of robustness. Indeed, if $A$ changes its battery energy allocation to $\Delta  E_{\mathrm{b},A} = \qty{-1}{\mega\joule}$, the lap time gain potential is significantly mitigated. One of the most favorable choice for $B$ loses $\qty{60}{\percent}$ of its potential, with a difference of $\qty{0.333}{\second}$. On the other hand, the choice $\Delta E_{\mathrm{b},B} = \qty{-1.4}{\mega\joule}$, is considerably more robust towards changes in the energy allocation of $A$, since the lap time gain potential is almost the same. The analysis of robust strategies can be extended to cover a large amount of typical situations arising during a race. For instance, favorable combinations of energy allocation between teammates can be chosen. 

\subsection{Lap Time Improvement Sensitivity}
\begin{figure}
\centering
\begin{externalize}{lapTimeVar_surf}
\input{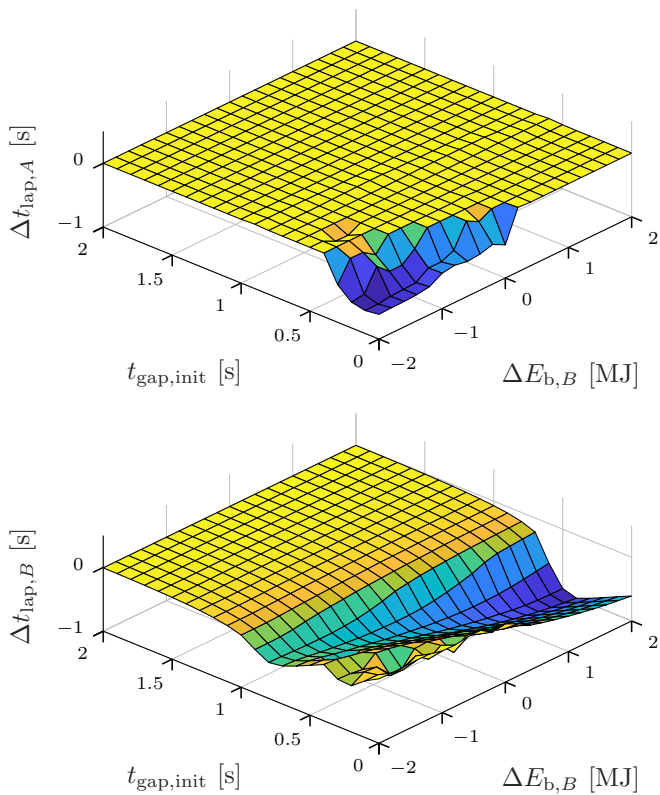}
\end{externalize}
\setlength{\abovecaptionskip}{\myfigskip}
\caption{Lap time improvement of $A$ (top) and $B$ (bottom) as a function of initial gap time and $B$'s allocated energy.}
\label{fig:lapTimeVar_surf}
\end{figure}

This last case study generalizes the previous one by means of sensitivity maps. Furthermore, it validates the robustness of the optimization framework over a large span of initial conditions, showing a clear and consistent trend. The initial conditions used are 
\begin{equation}
\BCvector = \begin{bmatrix} \qty{0}{\mega\joule} & \Delta\tilde E_{\mathrm{b},B} & \tilde t_\mathrm{gap,init}\end{bmatrix}^{\intercal},
\end{equation}
where $\Delta \tilde E_{\mathrm{b},B}\in [\qty{-2}{\mega\joule},\qty{2}{\mega\joule}]$ and $\tilde t_\mathrm{gap,init}\in[\qty{0}{\second},\qty{2}{\second}]$. The resulting $\Delta t_{\mathrm{lap},i}$ are shown in \Cref{fig:lapTimeVar_surf}. \\
The flat regions shows the conditions for which an agent has no influence on the other agent. They are clearly distinguishable since they show no improvement in lap time, and the agent behaves as if it were alone on the track. When the flat regions coincide, the interaction between agents is completely absent. This is the case for combinations of large initial gap times and lower energy budget for $B$. Where $B$ does not overtake, $A$ cannot profit at any time from the wake of $B$ and it is not influenced by its presence. In these cases, the surface of $A$ remains flat, although $B$ improves its lap time.

The exploitation of the drag reduction by $B$ is clearly visible in its surface's inflection. The shape is mostly influenced by the choice of the drag reduction model, and the maximum exploitation of the wake leads to a $\Delta t_{\mathrm{lap},B}= \qty{-0.994}{\second}$, assuming an optimal energy management. Typical allocations during a race are, for instance, $\Delta E_{\mathrm{b},B} =\Delta E_{\mathrm{b},A}= \qty{0}{\mega\joule}$. For this case, $A$ is not overtaken by $B$ and it cannot profit from the wake effect. To retain the lap time optimal, $A$ can thus employ a standard energy management. On the other hand, $B$ can switch to the game-theoretic energy management reducing the gap time, and potentially overtaking on the next lap. 

The region where $A$ improves its lap time corresponds to a successful overtake by $B$. This is the case for initial gap times smaller than $\qty{0.5}{\second}$ and a sufficient energy budget of $B$. The same region in the surface of $B$ shows an inverted trend, where its lap time gain is reduced compared to the bottom of the surface's inflection. As soon as $B$ overtakes and is in front, it cannot profit from the potential drag reduction for the remainder of the lap. The sooner the overtake occurs, the more the lap time of $B$ approaches its free-stream value. This effect is enhanced for small initial gap times and generous energy budgets for $B$.

With the developed framework, maps can be crafted to investigate if the interaction is relevant enough to be exploited, and if it is beneficial to switch from a standard energy management strategy to a game-theoretic one. Additionally, knowing if an overtake is possible is extremely relevant, and even though $B$ cannot always overtake, these results are helpful to develop strategies over subsequent laps to reduce the gap time. 

\section{Conclusion and Outlook}\label{sec:conclusion}
In this paper we have presented an optimization framework which includes two \gls{f1} cars on a single lap. Their interaction is captured by a drag reduction model which emulates the wake effect. To assess its impact on the energy management strategy, we have chosen a game-theoretic approach. The decision-making process arising from the interaction is captured by a Stackelberg game, which is mathematically described as a bilevel program. To this end, we have formulated the two-car-problem in this form. By means of the \gls{kkt} conditions, the structure of a single-level \gls{nlp} is recovered, allowing for a computationally efficient framework of the large-scale problem. 

This approach highlights the importance of including relevant interactions in racing problems, because the lap time improvement potential is considerable. First, we investigated an overtake scenario. The physical validity of the drag reduction model was assessed, showing that the overtake takes place at the end of a straight, as in a real race. Comparing the obtained solution with a benchmark, a new energy management strategy emerged. The saved drag energy is unevenly redistributed over the lap, unlike the classical strategy where only one agent is considered. Second, we discussed the impact of the energy allocation between agents. By varying the initial gap time, it is possible to predict if an overtake is possible and where to execute it. By changing the energy allocated by the other agent, it allows to identify allocation strategies which are more robust in terms of lap time gain. Finally, we identified the regions where the interaction becomes relevant within a lap, and we showed the validation of the framework for a large number of initial conditions. With the employed assumptions, the lap time gain potential is shown to be in the order of the tenth of seconds over one lap, a crucial advantage in the competitive sport of \gls{f1} racing.

The presented work serves as a basis to develop new energy management strategies by considering the presence of other agents. The framework can potentially accommodate further relevant interactions, such as the reduction in downforce or the use of the \gls{drs}. Including these models will contribute towards realistic implementations of game-theoretic strategies in real-world racing, allowing for comparison with current F1 energy management methods.

Online control strategies can also be derived, for instance, by formulating a game-theoretic \gls{mpc} for the energy management. By using a finite horizon, predictions about the future possible states of the opponent could be extrapolated and used for planning. However, the question about who is the leader and the follower is essential when employing these algorithms, as studied in \cite{cinar2024does}. Another approach is to explore the possibility to extend the dimensionality of the power split maps of \cite{salazar2017time,salazar2017real} by adding the gap time as an input. 

To extend the applicability on real racing conditions, human-like behavior could be introduced via sequential simulations in combination with learning-based methods. For instance, the energy management policies may be embedded in a reinforcement learning environment, where the agents represent the pilots. Receding horizon approaches already mimic the human logic of observe--plan--act, and incorporating uncertainty in the opponent's cost function could improve the robustness towards unexpected moves at the expense of optimality. Unlike the non-causal optimization presented in this work, sequential simulations allow the inclusion of disturbances, resulting in causal solutions.

Furthermore, the reformulated Stackelberg game allows for a \gls{svo} model of the cost function. The latter can be used to investigate combinations of competitive and prosocial behaviors, mainly for \gls{mpc} or reinforcement learning applications.

\section*{Acknowledgments}
We thank Ferrari S.p.A. for supporting this project. Moreover, we would like to express our deep gratitude to Ilse New for her helpful and valuable comments during the proofreading phase. We also appreciate the feedback provided by Fabio Widmer and Stijn van Dooren on earlier versions of this article.

\bibliographystyle{IEEEtran}
\bibliography{bibliography.bib}

\begin{IEEEbiography}[{\includegraphics[width=1in,height=1.25in,clip,keepaspectratio]{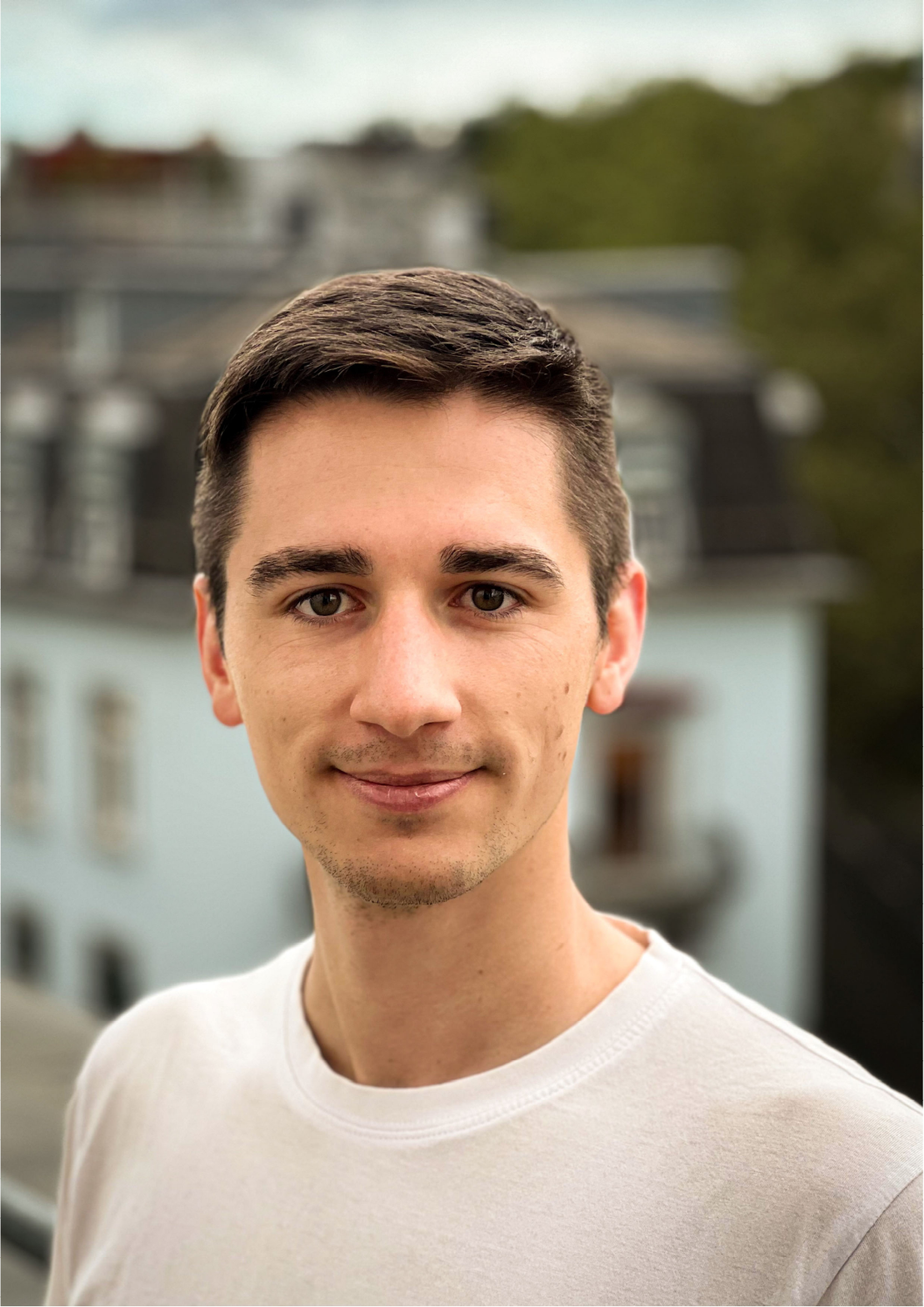}}]{Giona Fieni} was born in Mendrisio, Switzerland, and grew up near Lugano. He received his BSc and MSc degree in mechanical engineering from ETH Zürich in 2018 and 2021, respectively. Since November 2022 he is enrolled as Doctoral Student. His research fields include engine systems, race car power unit optimal control and multi-agent dynamical systems. He focuses on optimal control theory, game theory and reinforcement learning, mainly applied to hybrid electric vehicles.
\end{IEEEbiography}

\begin{IEEEbiography}[{\includegraphics[width=1in,height=1.25in,clip,keepaspectratio]{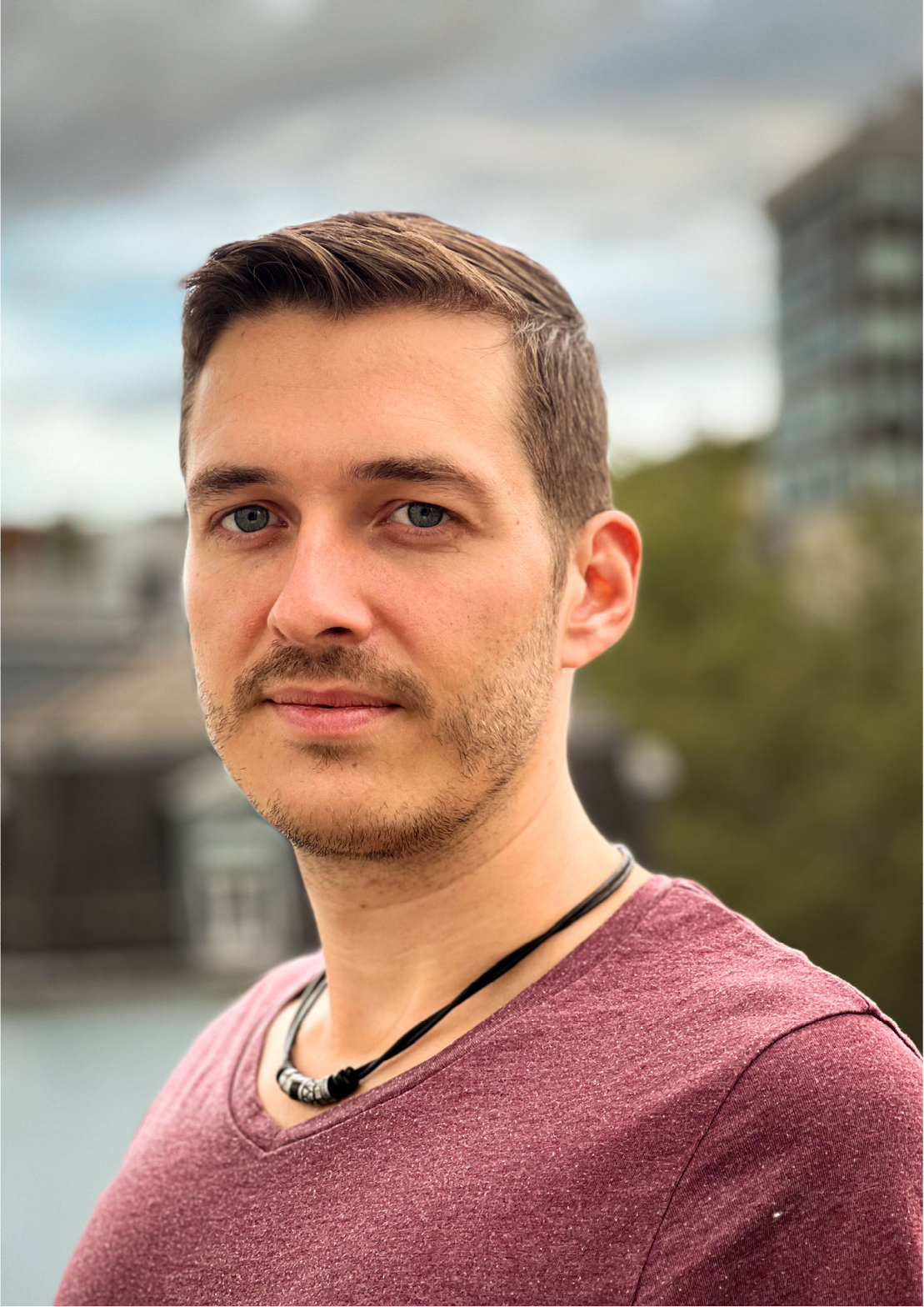}}]{Marc-Philippe Neumann}
	was born in Ludwigshafen am Rhein, Germany, and grew up near Lugano, Switzerland.
	He received his B.Sc. and M.Sc. degree in mechanical engineering from ETH Zürich in 2016 and 2019, respectively.
	Since July 2021 he pursues the Ph.D. degree with the Institute for Dynamic Systems and Control at ETH Zürich.
	His research focuses on hybrid electric race car powertrains, as well as optimal control theory and model predictive control.
\end{IEEEbiography}

\begin{IEEEbiography}[{\includegraphics[width=1in,height=1.25in,clip,keepaspectratio]{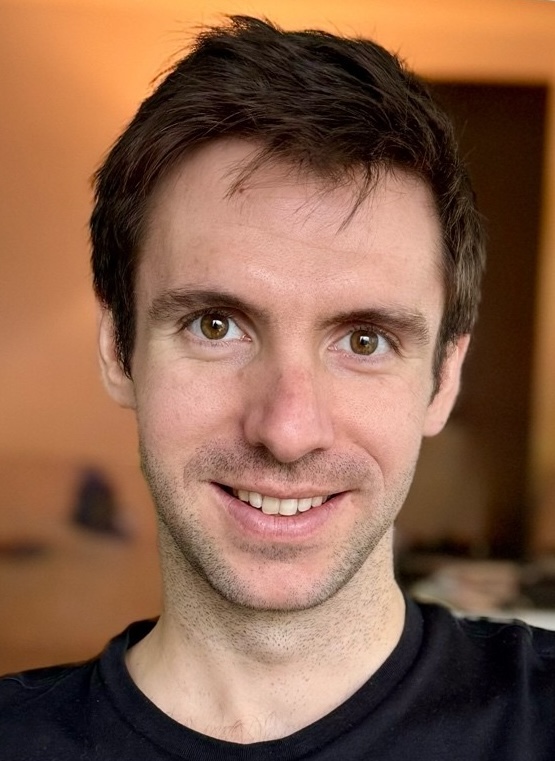}}]{Alessandro Zanardi} received his B.Sc. from Politecnico di Milano and the M.Sc. degrees in Robotics, Systems, and Control from ETH Zürich in 2015 and 2017, respectively. He has been a visiting researcher at the ABB Corporate Research in 2017, and at MIT Lids in Prof. Karaman’s group in 2023. He successfully defended his doctoral thesis at ETH in November 2023 pioneering the idea of "Urban Driving Games". Currently, he is employed at Embotech. His main research area involve motion planning, autonomous driving, machine learning.
\end{IEEEbiography}

\begin{IEEEbiography}[{\includegraphics[width=1in,height=1.25in,clip,keepaspectratio]{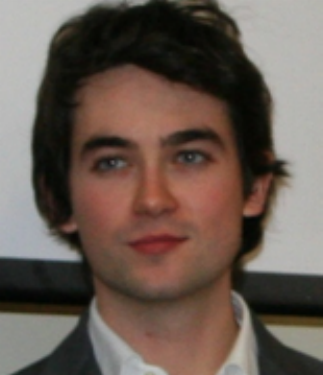}}]{Alberto Cerofolini}
	received the M.Sc. degree in Mechanical Engineering and the Ph.D. degree in Mechanics and Engineering Advanced Science from Universit\`{a} di Bologna, Italy, in 2009 and 2014, respectively. He currently holds a position as Power Unit Performance Engineer with the Power Unit Performance and Control Strategies Group of the Formula 1 team Scuderia Ferrari. His research focuses on lap-time-optimal and robust control strategies for the energy management of the Formula 1 car.
\end{IEEEbiography}

\begin{IEEEbiography}[{\includegraphics[width=1in,height=1.25in,clip,keepaspectratio]{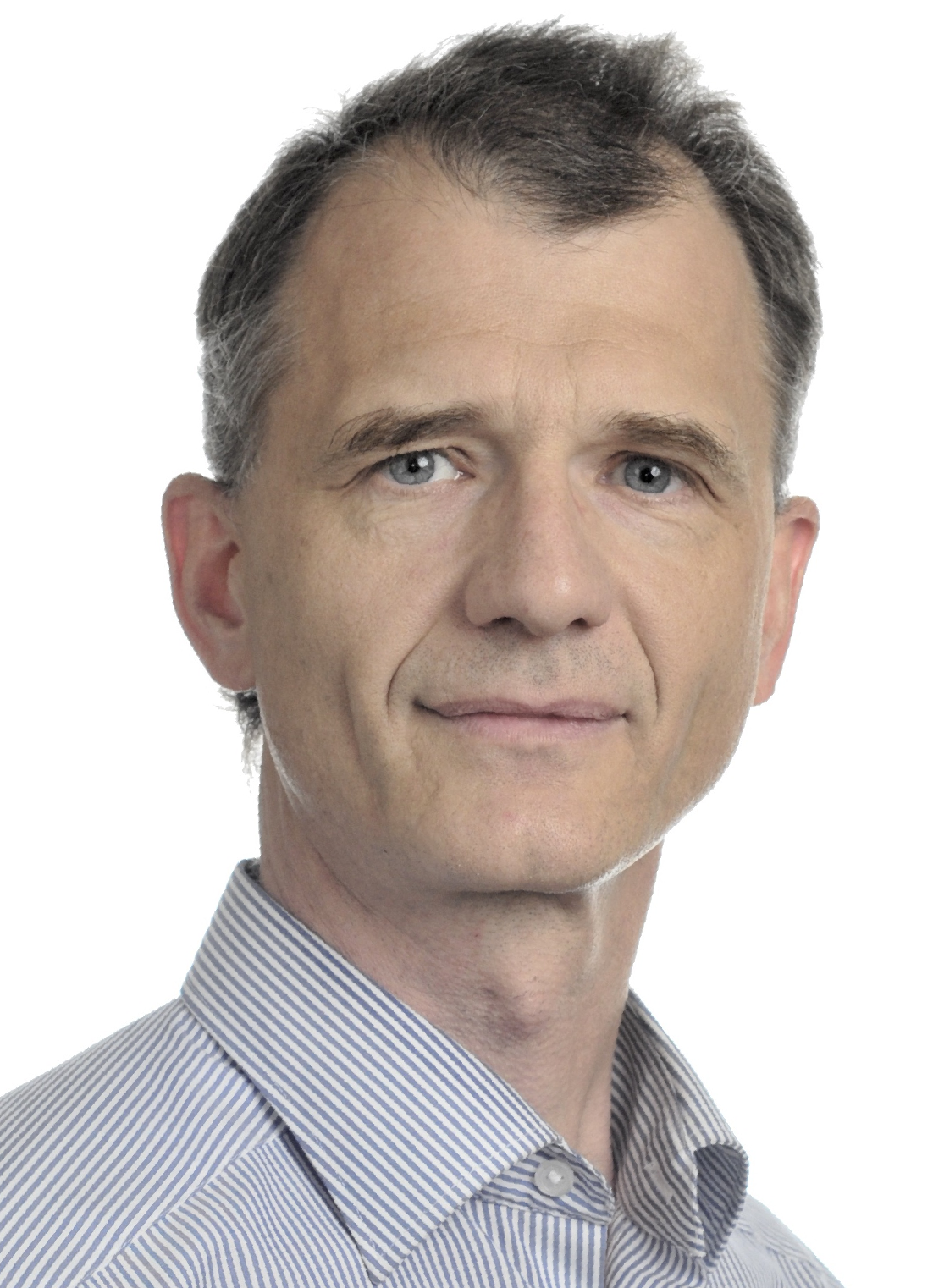}}]{Christopher H. Onder}
	received the Diploma and Ph.D.\ degrees in Mechanical Engineering from ETH Z\"urich, Switzerland. He is currently a Professor with the Institute for Dynamic Systems and Controls, ETH Z\"urich. He has authored or coauthored numerous articles and a book on modeling and control of engine systems. Prof.\ Dr.\ Onder was the recipient of the BMW Scientific Award, ETH Medal, Vincent Bendix Award, and Watt d'Or Energy Prize.
\end{IEEEbiography}

\vfill

\end{document}